\documentclass[prd,,aps,showpacs,nofootinbib]{revtex4}
\usepackage{amsfonts}
\usepackage{amsmath}
\usepackage{amssymb,epsfig}
\usepackage{mathrsfs}
\usepackage[usenames,dvipsnames]{color}
\usepackage[bookmarks]{hyperref}
\usepackage[usenames,dvipsnames]{color}
\usepackage{soul}

\newcommand{\Real}{\mathbb{R}}

\newcommand{\proofof}[1]{\noindent {\bf Proof of #1. }}
\newcommand{\qed}{\hfill \fbox{} \vspace{.3cm}}

\newtheorem{lemma}{Lemma}
\newtheorem{proposition}{Proposition}

\newtheorem{theorem}{Theorem}

\begin{document}

\title{Spherical accretion of a collisionless kinetic gas into a generic static black hole}
\author{Mehrab Momennia\footnote{Corresponding Author}}
\email{momennia1988@gmail.com}
\author{Olivier Sarbach}
\email{olivier.sarbach@umich.mx}
\affiliation{$^1$Instituto de F\'isica y Matem\'aticas, Universidad Michoacana de San
Nicol\'as de Hidalgo, Edificio C-3, Ciudad Universitaria, 58040 Morelia,
Michoac\'an, M\'exico.}

\begin{abstract}
We present a nontrivial extension of the problem of spherical accretion of a collisionless kinetic gas into the standard Schwarzschild black hole. This extension consists of replacing the Schwarzschild black hole by generic static and spherically symmetric black hole spacetimes with the aim of studying the effects of either modified gravitational theories beyond Einstein gravity or matter sources coupled to general relativity on the accretion process. This generalization also allows us to investigate the accretion into other types of black hole spacetimes, such as ones inspired by loop quantum gravity and string theory. To do so, we take into account a large class of static and spherically symmetric black holes whose spacetime is asymptotically flat with a positive total mass, has a regular Killing horizon, and satisfies appropriate monotonicity conditions of the metric functions. We provide the most general solution of the collisionless Boltzmann equation on such spacetimes by expressing the one-particle distribution function in terms of suitable symplectic coordinates on the cotangent bundle, and we calculate the relevant observables, such as particle current density and energy-momentum-stress tensor. Specializing to the case where the gas is described by an isotropic ideal fluid at rest at infinity, we compute the mass accretion rate and compression ratio, and we show that the tangential pressure is larger than the radial one at the horizon, indicating that the behavior of a collisionless gas is different from the one of an isotropic perfect fluid. Our general relations for the spacetime observables are given in terms of the generic metric functions which are determined by the parameters that characterize the black hole. As a concrete example, we apply our generic formulae to two special black hole spacetimes, namely the Reissner-Nordstr\"om black hole and a loop quantum corrected black hole. We explore the effects of the free parameters on the observables and accretion rate, and we compare the results with those corresponding to the standard Schwarzschild black hole.
\end{abstract}

\date{\today}
\maketitle


\section{Introduction}


Relativistic kinetic theory describes the dynamics of a dilute gas propagating on a curved spacetime. This theory allows one to study a wide
range of macroscopic astrophysical and cosmological phenomena from the
microscopic laws of particle mechanics, such as observations of supermassive
black holes hosted at the center of galaxies at scales smaller than their
gravitational radius \cite{EHT}, the description of stellar distribution in
the black hole environment \cite{pP72,jBrW76,sSsT86,slShapiro2023DM}, the modeling of dark
matter \cite{sCtSsPgPyS15,slShapiro2023DM}, probing the early Universe \cite%
{cMeB95,dBgDuHmMjN16,dBgDuHmMjN16b}, and, in the nonlrelativistic regime, the modeling of a galaxy as a
self-gravitating kinetic gas \cite{BinneyTremaine-Book}. In addition, the solutions to the collisionless Boltzmann equation in black hole spacetime backgrounds have been investigated for the accretion problem \cite{pRoS16,pRoS17,aCpM20,pMoA21a,pMoA21b,aGetal21,aCpMaO2022,pMaCaO2023,aCpMaO2024,gKpM2025}, for modeling stationary disks \cite{cGoS22a,cGoS22c} (see also \cite{fJ21,fJ22} for the self-gravitating case), for the mixing of a gas consisting of massive particles~\cite{pRoS18,pRoS20,pRoS24}, and decay estimates for the energy-momentum-stress tensor in the massless~\cite{lApBjS18,lB20,lBrR25} and massive~\cite{rR24} particles cases. Therefore, relativistic kinetic theory helps to gain deeper insights into complicated phenomena in extreme astrophysical environments by exploring the microscopic laws of particle motion in curved spacetimes and the collective behavior of the probe particles. In this theory, the state of the gas is characterized by the one-particle distribution function which is a solution of the relativistic Boltzmann equation. The macroscopic observables, such as particle density, energy density, pressure, mean particle velocity, etc. can be calculated from suitable fiber integrals of the distribution function over the momentum space.

In general, the dynamics of a kinetic gas requires solving the coupled system consisting of the Einstein-Maxwell-Boltzmann equations when the self-gravity is important and the particles are charged. There have been a lot of interesting mathematical results on this system in recent years. In the collisionless case, the global nonlinear stability of Minkowski spacetime~\cite{mT16,dFjJjS21}, the future stability of the Universe~\cite{Ringstrom-Book,hAhR16,dF16,hBdF22}, and the complete
gravitational collapse of a spherical cloud \cite{aRjV10,hAgR10,hAmKgR11b,hA12,hA14,aAmC14,hAgR25}
have been studied~(see \cite{hA11} for a review). Furthermore, the relativistic Boltzmann equation including
the collision term has been analyzed~\cite{dByC73,pNnNaR04,hLaR13}. Over the
years, the mathematical structure of the relativistic kinetic theory was
built by employing covariant formulations on the tangent or cotangent bundle associated
with the spacetime manifold and taking advantage of the bundle metric and
symplectic structure \cite{rL66,jE71,jE73,oStZ13,oStZ14a,oStZ14b} (for a recent review and references to further applications of the theory, see Ref.~\cite{rAcGoS22}).

Recent observations of supermassive black holes' shadow \cite{EHT} and detections of gravitational waves from stellar mass black holes \cite{LIGO} have indicated significant uncertainties in estimating the angular momentum of black holes. The uncertainty in estimating the black hole parameters leaves room for the existence of black hole solutions in gravitational theories beyond Einstein vacuum gravity. In this regard, black hole solutions in modified theories of gravitation, in Einstein gravity coupled to various matter sources, and even other types of black hole spacetimes inspired by loop quantum gravity (LQG) and string theory have free parameters describing the properties of the underlying theory and its black hole configurations. 
The footprint of these free parameters is encoded in the relevant observables of the accretion into black holes, such as current density and energy-momentum-stress tensor. In this article, we aim to generalize the work in~\cite{pRoS16} that describes the accretion of a collisionless kinetic gas on the standard Schwarzschild background to generic static and spherically symmetric black hole backgrounds. In particular, this extension helps to explore potential effects of the free parameters on the accretion process of black hole spacetimes beyond Einstein vacuum gravity. Although our study leaves out stationary rotating black holes, it provides a first step into this direction.

In this work, as in~\cite{pRoS16}, we assume that the gas is dilute enough such that its self-gravity and collisions can be ignored. Furthermore, we assume the accretion rate is so small that the spacetime background can be considered to be static, and if other matter fields are present, they do not interact directly with the gas particles. For simplicity, we restrict ourselves to identical gas particles of positive rest mass $m$. Therefore, we can model the gas by a solution of the collisionless Boltzmann equation on a fixed static background spacetime. In this article, we consider a large class of static and spherically symmetric  backgrounds which describe black hole spacetimes which are asymptotically flat, have positive total mass, a regular Killing horizon, and satisfy appropriate monotonicity conditions for their metric functions. We use the
same notation as in Refs.~\cite{pRoS16} and \cite{rAcGoS22}, except that we denote the
spacetime manifold by $\mathcal{M}$. Therefore, $(\mathcal{M},g)$ represents a
smooth, four-dimensional time-oriented Lorentzian manifold with the
signature convention $(-,+,+,+)$ for the metric. We follow the Hamiltonian formalism
and perform the computations on the cotangent bundle $T^{\ast }%
\mathcal{M}$ associated with the spacetime manifold $(\mathcal{M},g)$. We
employ the geometrized units $G=1=c$\ except in section~\ref{SubSec:SSS}
where the spherical steady-state accretion is investigated.

The outline of this paper is as follows. In the next section, we describe our model for the static and spherically symmetric black hole spacetimes and introduce horizon-penetrating coordinates that facilitate the interpretation of the spacetime observables at the horizon. Next, we explore the structure of the invariant subsets $\Gamma _{m,E,L_{z},L }\subset T^{\ast }\mathcal{M}$ defined by the conserved quantities $m$, $E$, $L_z$, and $L$ associated with the geodesic motion. To this purpose, we analyze the qualitative behavior of the generic effective potential related to the free test particles, and we separate the gas molecules into the two categories of absorbed and scattered particles. Next,
we discuss the phase flow and present the most general collisionless
distribution function on generic static backgrounds that is a solution to
the collisionless Boltzmann equation.

In section~\ref{Sec:Observables} we calculate the relevant spacetime observables, namely the particle current density and the energy-momentum-stress tensor, on the spacetime manifold $\mathcal{M}$.
Next, we consider the particular case of a distribution function depending only on the energy to further simplify the fiber integrals defining the observable.

Section~\ref{Sec:Accretion} is devoted to the application of the integrals obtained in section~\ref{Sec:Observables} to the spherical steady-state accretion process for a collisionless gas. In the asymptotic region, we assume the gas is described by distribution function depending only on the energy of the particles. The particle and energy fluxes $j_n$ and $j_{\varepsilon }$ corresponding to a sphere of constant radius outside the event horizon are computed, from which the mass and energy accretion rates are calculated. Further, we calculate the particle density $n_{\infty }$, energy density $\rho_\infty$, pressure $p_{\infty }$, and mean velocity of the particles at infinity and show that they behave as an isotropic ideal fluid. Next, we compute the particle density $n_{h}$, energy density $\rho _{h}$, radial pressure $p_{rad,h}$, and tangential pressure $p_{tan,h}$ of the gas on the horizon. We show that these quantities only depend on the horizon radius $r_h$ and integrals involving a function $L_c(E)$ of the energy which separates the absorbed and the scattered particles. Then, we consider the special case of the Maxwell-J\"uttner distribution function at infinity, and compute the mass and energy accretion rates, and the compression ratio at the horizon. We also calculate the particle density, energy density, and
the radial and tangential pressures at the event horizon and spatial infinity for the two special cases of low-temperature limit and ultra-relativistic limit at the horizon. Although the gas behaves like an isotropic perfect fluid at infinity, we explicitly show that, at least in the low and high temperature limits, this is not the case on the horizon due to the fact that the tangential pressure is larger than the radial one. Therefore, for a generic static black hole spacetime, the gas behaves differently than an isotropic perfect fluid on the horizon, just like it does in the Schwarzschild case. Next, we apply our general results for generic static and spherically symmetric spacetimes to two concrete examples: the Reissner-Nordstr\"om black hole and a loop quantum corrected black hole, and we investigate the effects of the electric charge and the deformation parameter on the observable quantities. Finally, we finish our paper with some concluding remarks in section~\ref{Sec:Conclusions}. Technical details are provided in the appendices.


\section{Most general solution of the collisionless Boltzmann equation on a generic static and spherically symmetric black hole background}

\label{Sec:Distribution} 

In this section, we first specify our model for the static, spherically symmetric black hole spacetimes and introduce horizon-penetrating coordinates. Next, we provide a detailed discussion of the phase flow and construct the most general solution of the collisionless Boltzmann equation, following the procedure in Ref.~\cite{pRoS16}.

\subsection{Generic black hole model and horizon-penetrating coordinates}
\label{SubSec:GenericBH}

As mentioned in the introduction, we are interested in studying the accretion process in
asymptotically flat black hole background spacetimes described by a generic static and spherically symmetric metric. In standard Schwarzschild coordinates $(\bar{t}%
,r,\vartheta ,\varphi )$, the metric of a static and spherically symmetric black
hole is given by 
\begin{equation}
g=-\chi \left( r\right) d\bar{t}\otimes d\bar{t} + \frac{1}{%
\psi (r)}dr\otimes dr+r^{2}\left( d\vartheta {\footnotesize %
\otimes }d\vartheta +\sin ^{2}\vartheta \,d\varphi \otimes %
d\varphi \right) ,  \label{metric}
\end{equation}%
with $\chi \left( r\right) =\sigma ^{2}\left( r\right) \psi
(r)$ where $\sigma \left( r\right) $ and $\psi (r)$ are smooth functions of the radial coordinate $r$ only. Consequently, the spacetime background described by the metric~(\ref{metric}) is invariant under the flow of the
Killing vector field $\mathbf{k}:=\partial _{\bar{t}}$ and with respect to rotations on
the $2$-spheres of constant $\bar{t}$ and $r$. Note that the functions $\chi$ and $\psi$ possess geometric interpretations: $-\chi = g(\mathbf{k},\mathbf{k})$ is the square norm of the Killing vector field $\mathbf{k}$ and $\psi = g^{-1}(dr,dr)$ is the square norm of the differential of the areal radius $r$. For the Schwarzschild metric, $\sigma \left( r\right)
=1$ and $\psi (r)=1-2M/r$, where $M$ is the black hole mass, and the event horizon is located at $r=2M$. 

However, in this article we consider a more general class of static and spherically symmetric black holes for which the metric functions $\sigma ,\psi ,\chi :\left( 0,\infty \right) \rightarrow \mathbb{R}$ are smooth and satisfy the following properties \cite{eCoS15a}: $\chi = \sigma^2\psi$ and
\begin{enumerate}
\item[(i)] Asymptotic flatness with positive total mass $M$: $\sigma
\rightarrow 1$, $\chi \rightarrow 1$, and $r^{2}\chi
^{\prime }\left( r\right) \rightarrow 2M>0$ for $r\rightarrow \infty $. Here and in what follows, a prime denotes a derivative with respect to the radial coordinate $r$.

\item[(ii)] Regular Killing horizon: $\sigma \left( r_{h}\right) >0$
and $\chi ^{\prime }\left( r_{h}\right) >0$ at the outermost horizon $%
r_{h}>0$ defined as the largest root of $\chi$.

\item[(iii)] Monotonicity conditions: $\sigma \left( r\right) >0$, $%
\chi \left( r\right) >0$, and $\chi ^{\prime }\left( r\right) >0$ for all $%
r>r_{h}$.

\item[(iv)] Single marginally stable circular orbit (MSCO) condition:
the function
\begin{equation}
\frac{\chi ^{\prime \prime}(r)}{\chi^\prime(r)}-2\frac{\chi^\prime(r)}{\chi(r)}+\frac{3}{r}
\label{Eq:SingleISCO}
\end{equation}
should have a single zero in the interval $r \in (r_{h},\infty)$. This zero gives the radius of the innermost stable circular orbit (ISCO).
\end{enumerate}
It is worth mentioning that the first two conditions are satisfied for any static, spherically symmetric, and asymptotically flat black hole spacetimes with total mass $M>0$ and non-degenerate event horizon at $r=r_{h}$. Whereas we have found that the third and fourth conditions are satisfied for all the examples considered in Section~\ref{SubSec:Examples}, it can be violated in extreme conditions in which the black hole differs much from the Schwarzschild one. For instance, the fourth condition is violated in black hole spacetimes containing multiple photon spheres outside the event horizon~\cite{MultiPhSphBHs}. Finally, as we will prove further below, the zeroes of the function defined in Eq.~(\ref{Eq:SingleISCO}) are inflection points of the effective potential for the radial motion and describe MSCOs. Conditions (i)--(iii) guarantee that there is at least one such MSCO (see Lemma \ref{Lem:Potential} below); hence (iv) demands that such MSCOs are unique, and thus correspond to an ISCO. Furthermore, as we will see, conditions (i)--(iv) imply the existence of a single photon sphere outside the horizon.

In this regard, note that the $4$-wave vector $k$ of a photon in a generic static spacetime of the form~(\ref{metric}) must satisfy $g_{\mu \nu }k^{\mu }k^{\nu }=0$ which leads to~\cite{dMmMaH2024}
\begin{equation}
(\sigma(r) k^{r})^{2}
 + L_\gamma^2\frac{\chi(r)}{r^2} = E_\gamma^2,
\label{gEL}
\end{equation}
where $E_{\gamma }$ and $L_{\gamma }$ are the energy and total angular momentum of the photons, respectively. Therefore, circular photon orbits correspond to critical points of $\chi(r)/r^2$ and hence one obtains the following relation describing the photon sphere radius $r_{ph}$
\begin{equation}
2\chi(r_{ph})=r_{ph}\chi^{\prime}(r_{ph}).
\label{Eq:PhotonSphere}
\end{equation}

Since we are interested in computing the observables associated with the
kinetic gas at the future horizon and not just outside it, we replace the Schwarzschild coordinates with the regular coordinates $(t,r,\vartheta ,\varphi )$. Here, the new time coordinate $t$ is related to the Schwarzschild time $\bar{t}$ as follows:
\begin{equation}
d\bar{t}=dt-\frac{dr}{\sigma \left( r\right) \psi (r)}+\eta (r)dr,
\label{Eq:ttbar}
\end{equation}
with a smooth function $\eta :(0,\infty )\rightarrow \mathbb{R}$. Under this
coordinate transformation, the metric (\ref{metric}) reads
\begin{equation}
g=-\chi dt\otimes dt + 2(\sigma -\chi \eta )dt \otimes_s dr + \eta (2\sigma -\chi \eta )dr\otimes dr+r^{2}\left(
d\vartheta \otimes d\vartheta +\sin ^{2}\vartheta \,d\varphi 
\otimes d\varphi \right) ,
\end{equation}%
which is free of coordinate singularities ($\det(g)=- \sigma^2 r^4 \sin^2 \vartheta$) for $r>0$ except at the poles $%
\vartheta =0,\pi $.\footnote{These coordinate singularities could be removed by replacing $(r,\vartheta
,\varphi )$ with Cartesian-like coordinates $(x,y,z)=r(\cos \varphi \sin
\vartheta ,\sin \varphi \sin \vartheta ,\cos \vartheta )$. However, this will not be necessary since for simplicity, we will remove the (zero measure set) consisting of polar orbits.} Here and in the following, $\otimes_s$ denotes the symmetrized tensor product, defined as $v\otimes_sw:=(v\otimes w+w\otimes v)/2$. The smooth
function $\eta $ describes the freedom in choosing the foliation of the
black hole spacetime which is regular at the future horizon. The constant time $t = const$ surfaces are spacelike for $\eta (2\sigma -\chi \eta ) > 0$ while they
are incoming null surfaces for $\eta =0$. In the latter case, the corresponding coordinates $(t,r,\vartheta ,\varphi )$ generalize the  ingoing Eddington-Finkelstein coordinates (see \cite{MTW-Book} for instance) to generic static spherically symmetric black holes.

Henceforth, we shall choose $\eta=1$ such that the inverse metric has the following explicit form
\begin{equation}
g^{-1}=-\frac{2-\sigma \left( r\right) \psi (r)}{\sigma \left( r\right) }%
\partial _{t}\otimes \partial _{t}+2\frac{1-\sigma \left( r\right) \psi (r)}{%
\sigma \left( r\right) }\partial _{t}\otimes _{s}\partial _{r}+\psi
(r)\partial _{r}\otimes \partial _{r}+\frac{1}{r^{2}}\left( \partial
_{\vartheta }\otimes \partial _{\vartheta }+\frac{1}{\sin ^{2}\vartheta }%
\partial _{\varphi }\otimes \partial _{\varphi }\right).
\label{Eq:SchwarzschildInv}
\end{equation}%
Hereafter, we concentrate on the
region $\mathcal{M}$ of the (extended) black hole manifold which is covered by the horizon-penetrating coordinates $(t,r)$ with $t\in \mathbb{R}$ and $r\geq r_h$. Note that the $t = const$ surfaces are spacelike as long as $\sigma\psi < 2$, a condition that is always satisfied near the horizon and in the asymptotic regime according to conditions (i) and (ii). In any case, the condition $\sigma\psi < 2$ will not be relevant for the following discussion.

\subsection{Hamiltonian flow, conserved quantities, invariant
submanifolds, and effective potential}

The free-particle Hamiltonian calculated from the inverse metric~(%
\ref{Eq:SchwarzschildInv}) is given by 
\begin{equation}
H(x,p) := \frac{1}{2} g^{\mu\nu}(x) p_\mu p_\nu = \frac{1}{2}\left[ -\frac{2-\sigma \left( r\right) \psi (r)}{\sigma
\left( r\right) }p_{t}^{2}+2\frac{1-\sigma \left( r\right) \psi (r)}{\sigma
\left( r\right) }p_{t}p_{r}+\psi (r)p_{r}^{2}+\frac{1}{r^{2}}\left(
p_{\vartheta }^{2}+\frac{p_{\varphi }^{2}}{\sin ^{2}\vartheta }\right) %
\right],
\label{Eq:HSchwarzschild}
\end{equation}
and it is defined on the one-particle phase space
\begin{equation}
\Gamma:= \{ (x,p)\in T^{\ast }\mathcal{M}: \hbox{$p$ is future-directed timelike} \}.
\end{equation}

Besides, the following quantities are conserved along the particle
trajectories:
\begin{eqnarray}
m=\sqrt{-2H} &&\hbox{(rest mass)}, \\
E=-p_{t} &&\hbox{(energy)}, \\
L _{z}=p_{\varphi } &&\hbox{(azimutal angular momentum)}, \\
L =\sqrt{p_{\vartheta }^{2}+\frac{p_{\varphi }^{2}}{\sin ^{2}\vartheta }}
&&\hbox{(total angular momentum)},
\end{eqnarray}%
because the spacetime metric (\ref{metric}) is stationary and spherically
symmetric. Accordingly, we define the smooth functions $F_{0},F_{1},F_{2},F_{3}: \Gamma\rightarrow \mathbb{R}$ on the cotangent
bundle as follows
\begin{equation}
F_{0}(x,p):=-H(x,p),\qquad F_{1}(x,p):=-p_{t},\qquad F_{2}(x,p):=p_{\varphi
},\qquad F_{3}(x,p):=p_{\vartheta }^{2}+\frac{p_{\varphi }^{2}}{\sin
^{2}\vartheta },  \label{Eq:FalphaDef}
\end{equation}%
and it is easy to verify that these quantities Poisson-commute with each other,
\begin{equation}
\{F_{i},F_{j}\}=0,\qquad i,j=0,1,2,3.
\end{equation}

In addition, for each given value of the conserved quantities $(m,E,L _{z},L)$, we consider the (possibly empty) subset of the
one-particle phase space defined as
\begin{equation}
\Gamma_{m,E,L _{z},L }:=\left\{ (x,p)\in \Gamma :F_{0}(x,p)=\frac{1}{2%
}m^{2},F_{1}(x,p)=E,F_{2}(x,p)=L _{z},F_{3}(x,p)=L ^{2}\right\} ,
\end{equation}
which, by construction, is invariant under the Hamiltonian flow. The main
result of this subsection is given in the following proposition, which generalizes Proposition~2 in~\cite{pRoS16} to the case of our generic black hole model:

\begin{proposition}
\label{Prop:GammaProp} Suppose that $E>m>0$ and that $0<|L_{z}|<L $. In
addition, suppose that the condition $L \neq L_{c}(E)$ holds (see Eq.~(\ref{lCritCond}) below for the definition of $L _{c}(E)$). Then, $\Gamma_{m,E,L_{z},L }$ is a smooth, four-dimensional submanifold of $\Gamma$ which is invariant with respect to the Hamiltonian flows associated with $F_{0}$, $F_{1}$, $F_{2}$, and $F_{3}$. Furthermore, the restriction of the Poincar\'e one-form $\Theta = p_\mu dx^\mu$ on $\Gamma_{m,E,L_z,L}$ is closed.

Finally, for $L > L_c(E)$, $\Gamma_{m,E,L_z,L}$ has only one connected component which contains points with arbitrary large values of $r$ (describing scattered particles), whereas for $0 < L < L_c(E)$ there are two such components (one of them describing absorbed particles, the other one particles that are ejected from the white hole). These components\footnote{For polar orbits which have $L_z=0$ these components have topology $\mathbb{R}^2\times S^2$ instead.} have topology $\mathbb{R}^{2}\times S^{1}\times S^{1}$.
\end{proposition}

The proof of this proposition follows the same arguments as in~\cite{pRoS16} and is based on a detailed understanding of the properties of the effective potential associated to the radial motion. Since the latter is not known in explicit form, the proof is more involved than in the Schwarzschild case. For this reason, let us start with a brief outline of the proof.

In a first step, one separates the condition for a point $(x,p)$ to lie in $\Gamma_{m,E,L_z,L}$ into conditions for the conjugate pairs $(t,p_{t})$, $(r,p_{r})$, $(\vartheta ,p_{\vartheta })$, and $(\varphi ,p_{\varphi
})$. In a second step, one analyzes the properties of the polar motion. The next step consists in understanding the effective potential describing the radial motion (this is the step which takes most efforts). Finally, in step four, one completes the proof by showing that the differentials of $F_0$, $F_1$, $F_2$, and $F_3$ are everywhere linearly independent from each other provided $0 < |L_z| < L$ and $L \neq L_c(E)$.
\\

{\it Step 1:} Suppose $(x,p)\in \Gamma_{m,E,L_z,L}$. Then, the conjugate pairs $(t,p_{t})$, $(r,p_{r})$, $(\vartheta ,p_{\vartheta })$, and $(\varphi ,p_{\varphi
})$ satisfy the following constraints:
\begin{eqnarray}
(t,p_{t}) &:&p_{t}=-E,  \label{Eq:Ept} \\
(\varphi ,p_{\varphi }) &:&p_{\varphi }=L _{z}, \\
(\vartheta ,p_{\vartheta }) &:&p_{\vartheta }^{2}+\frac{L _{z}^{2}}{\sin
^{2}\vartheta }=L ^{2},  \label{Eq:thetaPlaneRestriction} \\
(r,p_{r}) &:&\psi (r)p_{r}^{2}-2\frac{1-\sigma \left( r\right) \psi (r)}{%
\sigma \left( r\right) }Ep_{r}-\frac{2-\sigma \left( r\right) \psi (r)}{%
\sigma \left( r\right) }E^{2}+m^{2}+\frac{L ^{2}}{r^{2}}=0.
\label{Eq:rPlaneRestriction}
\end{eqnarray}
These restrictions indicate that, whereas the time coordinate $t$ and the azimuthal angle $\varphi $ are free (giving rise to  $\mathbb{R}\times S^1$ in the topology of $\Gamma_{m,E,L_z,L}$), the pairs $(\vartheta ,p_{\vartheta })$ and $(r,p_{r})$ are restricted by Eqs.~(\ref{Eq:thetaPlaneRestriction}) and (\ref{Eq:rPlaneRestriction}), respectively. In the next two steps, we analyze the two-dimensional phase diagrams arising from these restrictions in detail.
\\

\begin{figure}[ht]
\centerline{\resizebox{8.5cm}{!}{\includegraphics{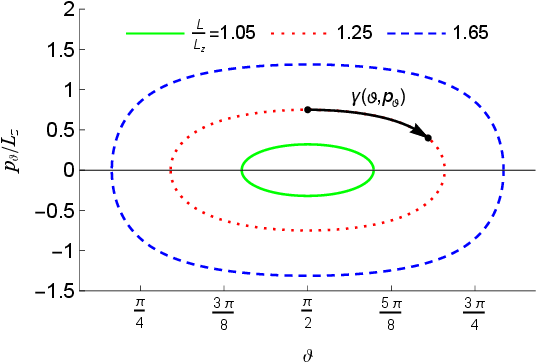}}}
\caption{Phase diagram illustrating the projection of the invariant
sets onto the $(\protect\vartheta,p_\protect\vartheta)$-plane for various
values of $L$. The curves shrink to a point in
the limit $L \to L_z$ which corresponds to motion confined to the
equatorial plane $\protect\vartheta = \protect\pi/2$.}
\label{Fig:plot_thpth}
\end{figure}

{\it Step 2:} As illustrated in figure~\ref{Fig:plot_thpth}, the pair $(\vartheta ,p_{\vartheta })$ is confined to closed curves winding around the equilibrium point $(\vartheta ,p_{\vartheta
})=(\pi /2,0)$ in the $(\vartheta ,p_{\vartheta })$-plane for $0<|L
_{z}|<L $, which has the topology of $S^1$. The limiting cases $L _{z}=0$ and $L_{z}=\pm L $ refer, respectively, to motion confined in a plane containing the $z$-axis and to motion in the equatorial plane.

{\it Step 3:} Next, we turn our attention to the set in the $(r,p_{r})$-plane defined by the constraint~(\ref{Eq:rPlaneRestriction}) in more detail. By multiplying both sides of the relation by $\chi (r)=\sigma ^{2}\left( r\right) \psi (r)$
and assuming $\psi (r)\neq 0$, one can represent this constraint as 
\begin{equation}
\sigma ^{2}\left( r\right) \left( \psi (r)p_{r}-\frac{1-\sigma \left(
r\right) \psi (r)}{\sigma \left( r\right) }E\right) ^{2}+V_{m,L
}(r)=E^{2},  \label{Eq:rPlaneRestrictionBis}
\end{equation}
where $V_{m,L }(r)$ is the effective potential which is defined as follows:
\begin{equation}
V_{m,L }(r):=\chi (r)\left( m^{2}+\frac{L ^{2}}{r^{2}}\right) ,
\label{Eq:Vmell}
\end{equation}%
in terms of the metric function $\chi (r)$ characterizing the properties of the spacetime background. According to
Hamilton's equations, the radial velocity of the particles is given by 
\begin{equation}
\dot{r}:=\frac{dr}{d\tau} = \frac{\partial H}{\partial p_{r}}=\psi
(r)p_{r}+\frac{1-\sigma \left( r\right) \psi (r)}{\sigma \left( r\right) }%
p_{t}=\psi (r)p_{r}-\frac{1-\sigma \left( r\right) \psi (r)}{\sigma \left(
r\right) }E,
\end{equation}
where $\tau$ denotes proper time. By comparing this relation with Eq.~(\ref{Eq:rPlaneRestrictionBis}),
we find that the set defined by the restriction~(\ref{Eq:rPlaneRestriction})
is equivalent to
\begin{equation}
\sigma ^{2}\left( r\right) \dot{r}^{2} + V_{m,L}(r)=E^{2},
\label{Eq:RadialVelocity}
\end{equation}
as long as $\psi (r)\neq 0$.

Solving Eq.~(\ref{Eq:rPlaneRestrictionBis}) for $p_r$, one obtains
\begin{equation}
p_{r}=p_{r\pm }(r):=\frac{\left[ 1-\sigma \left( r\right) \psi (r)\right]
E\pm \sqrt{E^{2}-V_{m,L }(r)}}{\sigma \left( r\right) \psi (r)}=\frac{%
\sigma \left( r\right) \left( m^{2}+\frac{L ^{2}}{r^{2}}\right) -\left[
2-\sigma \left( r\right) \psi (r)\right] E^{2}}{\left[ 1-\sigma \left(
r\right) \psi (r)\right] E\mp \sqrt{E^{2}-V_{m,L}(r)}},  
\label{Eq:prpm}
\end{equation}
where, according to Eq.~(\ref{Eq:RadialVelocity}), the $+$/$-$ sign characterizes the outgoing/incoming particles. Therefore, the phase space diagram in the $(r,p_r)$ plane can be understood by analyzing the structure of the effective potential $V_{m,L}$ and the phase curves for different energy levels $E$.

In the next lemma, we probe the qualitative behavior of the
effective potential for generic static and spherically symmetric spacetimes.

\begin{lemma}
\label{Lem:Potential} Let $E,m>0$ and $r\geq r_h$. In addition, let the norm $\chi (r)$ of the Killing vector field $\partial_{\bar{t}}$ satisfy the precise conditions (i)-(iv) described
after the metric (\ref{metric}), and define the smooth function $\mathcal{G}:[r_{h},\infty )\rightarrow \mathbb{R}$ according to
\begin{equation}
\mathcal{G}(r) :=\frac{2\chi (r)-r\chi ^{\prime }(r)}{r^{3}\chi
^{\prime }(r)}.
\label{Gfunction}
\end{equation}

Then, the function $\mathcal{G}$ has a unique root at $r_{ph}$ and a global maximum at $r_{ISCO}$, where $r_{ISCO} > r_{ph} > r_h$. Furthermore, depending on the value of $L$ compared to
\begin{equation}
L_{ISCO}:=\frac{m}{\sqrt{\mathcal{G}(r_{ISCO})}},
\end{equation}
the effective potential $V_{m,L}(r)$ behaves as follows;
\begin{enumerate}
\item[(a)] $V_{m,L}(r)$ is an increasing function of $r$ for $0 \leq L<L_{ISCO}$,

\item[(b)] $V_{m,L}(r)$ has an inflection point at $r = r_{ISCO}$ for $L=L_{ISCO}$,

\item[(c)] $V_{m,L}(r)$ has a local minimum and a local maximum for $L_{ISCO}<L < \infty$.
\end{enumerate}
\end{lemma}

\proofof{Lemma~\ref{Lem:Potential}} In order to prove this lemma, we first discuss the general behavior of the smooth function $\mathcal{G}$. By considering the restrictions (i)-(iii)
on the metric function $\chi \left( r\right) $, one finds that the function $\mathcal{G}\left( r\right) $ is well-defined and negative at the horizon $r_{h}$; in fact $\mathcal{G}\left( r_{h}\right) = -1/r_{h}^{2}$. On the
other hand, it falls off as $\mathcal{G}\left( r\right) $ $\simeq \left(
Mr\right) ^{-1}$ $\left( 1-3M/r\right) $ in the asymptotic region $r\rightarrow \infty $ which is positive valued. By taking these observations into account, one finds that $\mathcal{G}$ has a global maximum outside the horizon where $\mathcal{G} > 0$. By comparing the relations $\mathcal{G}^\prime (r)=0$ and $V_{m,L }^{\prime
}(r)=0=V_{m,L }^{\prime \prime }(r)$ obtained from Eqs.~(\ref{Gfunction}) and (\ref{Eq:Vmell}), respectively, one finds that both are equivalent to the zeroes of Eq.~(\ref{Eq:SingleISCO}). Thus, according to condition (iv), the maximum of $\mathcal{G}$ must be unique, and it is located at an inflection point of $V_{m,L}$ in the interval $(r_h,\infty)$. Such an inflection point describes an MSCO, and since it is unique, it corresponds to the ISCO.

\begin{figure}[ht]
\centerline{\resizebox{8.5cm}{!}{\includegraphics{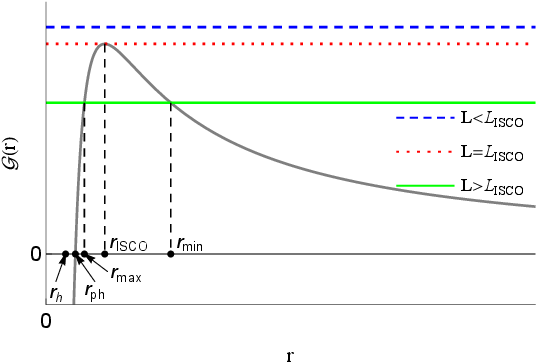}}}
\caption{Schematic illustration of the profile of $\mathcal{G}$ versus $r$. $\mathcal{G}$ has a unique root at $r_{ph}$ and a unique global
maximum at $r_{ISCO}$ outside the event horizon $r_{h}<r_{ph}<r_{ISCO}$. For $L >L_{ISCO}$, Eq. (\ref{DofV}) has two solutions that shrink to a single root in the
limit $L \rightarrow L _{ISCO}$.}
\label{GfunctionFig}
\end{figure}

Moreover, it follows from the previous arguments that $\mathcal{G}$ has a unique root at $r_{ph}\in (r_h,r_{ISCO})$. Since the numerator in Eq.~(\ref{Gfunction}) vanishes at $r_{ph}$, it describes the radius of the unique photon sphere outside the horizon, see Eq.~(\ref{Eq:PhotonSphere}). Figure~\ref{GfunctionFig} offers an illustration for the general behavior of the function $\mathcal{G}(r)$.

Having the qualitative behavior of the function $\mathcal{G}%
\left( r\right) $ at hand, we can describe the general qualitative behavior
of the effective potential as follows. First, from equation (\ref{Eq:Vmell}), we see that the potential
vanishes at the horizon $r_{h}$ whereas it is positive definite for large $r$
with the asymptotic behavior $V_{m,L }(r)$ $\simeq m^{2}$ $\left(
1-2M/r\right) $. In the case of $L =0$ given in $(a)$, $V_{m,L }(r)$ is increasing for all values of $r$ outside the event horizon, according to (iii). For $L \neq 0$, its
first derivative can be written as 
\begin{equation}
V_{m,L }^{\prime }(r)=L ^{2}\chi ^{\prime }(r)\left( \frac{m^{2}}{L
^{2}}-\mathcal{G}\left( r\right) \right) ,  \label{DofV}
\end{equation}
in terms of the function $\mathcal{G}\left( r\right) $. Now, there are three
possibilities based on the value of $m^{2}/L ^{2}$ (see Fig. \ref%
{GfunctionFig}). For $L >L _{ISCO}$, $V_{m,L }^{\prime }(r) = 0$ has two
solutions
\begin{equation}
r_{\min }\left( L \right) \text{ and }r_{\max }\left( L \right) ,
\label{GeneralRminRmax}
\end{equation}
corresponding to a local minimum and maximum of the effective potential on $%
[r_{h},\infty )$.\footnote{Note that the radii $r_{\min }\left( L \right) $ and $%
r_{\max }\left( L \right) $ satisfy the inequalities $r_{ph}<r_{\max
}\left( L \right) <r_{ISCO}<r_{\min }\left( L \right) $\ and they can
be obtained analytically/numerically once we have the explicit form of the
metric function.} Fig. \ref{GfunctionFig} shows that $r_{\max }\left( L
\right) $ decreases monotonically from $r_{ISCO}$ to $r_{ph}$ and $r_{\min
}\left( L \right) $ increases monotonically from $r_{ISCO}$ to $\infty $
when $L $ increases from $L _{ISCO}$ to $\infty $. This implies the behavior of the effective potential for $L_{ISCO} < L < \infty $ given in $(c)$.  

For $L =L _{ISCO}$ given in $(b)$, the two solutions $r_{\min }\left( L
\right) $ and $r_{\max }\left( L \right) $ of $V_{m,L }^{\prime }(r)$\
shrink to the single root $r_{ISCO}$\ which is an inflection point of $%
V_{m,L }(r)$. Finally, as for case $(a)$, $V_{m,L}(r)$\ is increasing for all $%
r>r_{h}$ when $0<L <L _{ISCO}$ and asymptotically approaches $m^2$.
\hfill \fbox{} 
\vspace{0.3cm}

{\bf Remark}: For the Schwarzschild black hole, one has $r_{h}=2M$, $r_{ph}=3M$, $%
r_{ISCO}=6M$, and $L _{ISCO}=\sqrt{12}Mm$.\\

It follows from Lemma~\ref{Lem:Potential} that the open range of parameters for having bound orbits is given by
\begin{equation}
L >L _{ISCO},\ \ \ \ \ \sqrt{V_{m,L }(r_{\min }\left( L \right) )
} < E < \min \left\{ \sqrt{V_{m,L }(r_{\max }\left( L \right) )}%
,m\right\} ,  \label{BoundOrbitCondition}
\end{equation}%
where the lower bound on $E$ corresponds to the stable circular orbit located at $r_{\min }\left( L \right)$.

However, for the accretion problem in this article, the unbound orbits are relevant. To further classify them, we first note that the the extrema of the effective potential, that is, $V_{m,L ,\max }:=V_{m,L
}(r_{\max }\left( L \right) )$\ and $V_{m,L ,\min }:=V_{m,L
}(r_{\min }\left( L \right) )$ are monotonically increasing with $L$. This follows from observing that
\begin{equation}
\frac{dV_{m,L ,\max}}{dL }=\chi (r_{\max }\left( L
\right) )\frac{2L }{r_{\max }^2\left( L \right) } > 0,
\end{equation}
and similarly for $V_{m,L ,\min }$. Therefore, as $L$ increases from $L_{ISCO}$ to $\infty$, $V_{m,L ,\max}$ increases monotonically from $E_{ISCO}^2:=V_{m,L_{ISCO}}(r_{ISCO})$ to $\infty$ and $V_{m,L ,\min}$ from $E_{ISCO}^2$ to $m^2$, as follows from Eq.~(\ref{Eq:Vmell}). In particular, given $E > E_{ISCO}$, there is a unique critical value of the total angular momentum$\ $%
given by $L =L _{c}\left( E\right) $\ for which the energy of the test
particle coincides with the maximum value of the effective potential, such that
\begin{equation}
E^{2}=V_{m,L _{c}(E),\max }
.\label{lCritCond}
\end{equation}
For the Schwarzschild and Reissner-Nordstr\"{o}m cases, $L_c(E)$ can be computed analytically~\cite{pRoS16,aCpM20}; however for more general black holes one might need to find $L_c(E)$ by finding $r_{max}(L)$ and solving Eq.~(\ref{lCritCond}) numerically. It follows from Eq.~(\ref{lCritCond}) that $L_c(E)\to \infty$ monotonically as $E\to \infty$, and since $r_{max}(L)\to r_{ph}$ in this limit, it follows from Eq.~(\ref{Eq:Vmell}) that
\begin{equation}
\tilde{\zeta} := \lim\limits_{E\to\infty}\frac{L_c(E)}{E} = \frac{r_{ph}}{\sqrt{\chi(r_{ph})}}.
\label{Eq:zetatilde}
\end{equation}
For instance, $\tilde{\zeta} = 3\sqrt{3} M$ for the standard Schwarzschild black hole, which is equal to the critical angular momentum for photons, as it should be. This limit will turn out to be relevant when discussing the ultra-relativistic limit of our accretion models in Section~\ref{SubSec:SSS}.

\begin{figure}[ht]
\centerline{\resizebox{8.5cm}{!}{\includegraphics{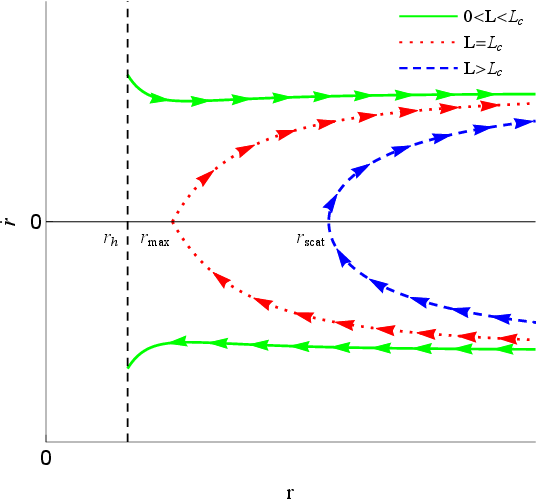}}}
\caption{Schematic illustration of the radial velocity $\dot{r}$ versus $r$ for different values of the angular momentum $L$. The continuous green curve with negative $\dot{r}$ shows a particle coming from infinity which is absorbed by the black hole (case (I)), whereas the continuous green curve with positive $\dot{r}$ corresponds to a particle emitted from the white hole and escapes to infinity. The dashed blue curve in the region $r > r_{ISCO}$ denotes an incoming particle from infinity that has large enough angular momentum $L > L_c(E)$ to be reflected at the potential barrier (case (II)). The dotted red curve corresponds to the separatrix and shows the energy level $E$ for which $L = L_c(E) $.}
\label{radialVelocityFigGeneral}
\end{figure}

Taking into account the above arguments and the radial velocity relation
\begin{equation}
\dot{r}=\dot{r}_{\pm }:=\pm \frac{\sqrt{E^{2}-V_{m,L }(r)}}{\sigma \left(
r\right) },  \label{radialVelocity}
\end{equation}%
we can distinguish between the two following cases:

\begin{enumerate}
\item[(I)] \emph{Absorbed particles $0<L <L _{c}(E)$.} In this
case, the radial velocity of particles (\ref{radialVelocity}) is regular at
the horizon with the value $\left. \dot{r}_{\pm }\right\vert _{r\rightarrow
r_{h}} = \pm E/\sigma \left( r_{h}\right) $, and behaves as $\dot{r}_{\pm }\approx \pm \sqrt{E^{2}-m^{2}}$ in the asymptotic region $r\rightarrow \infty $. Therefore,
the minus sign corresponds to trajectories of particles
coming from spatial infinity that are being absorbed by the black hole, whereas the positive sign describes particles emanating from the white hole that escape to infinity (see the continuous green curves in Fig.~\ref{radialVelocityFigGeneral} for a general illustration of the
trajectories). In this paper, we consider only the particles that are being absorbed by the
black hole and discard the particles that emanate from the white hole, since we are interested in astrophysical black holes.

\item[(II)] \emph{Scattered particles $L >L _{c}(E)$.} When the total angular momentum is higher than the critical value, the absolute value of $\dot{r}_{\pm}$ is lower than in the previous case (for fixed $E$ and $r$) since the value of $L$ is larger. Moreover, when $r$ decreases from infinity to the minimum of the effective potential, $\left\vert \dot{r}_{-}\right\vert$ slightly increases (not visible in the schematic representation of Fig.~\ref{radialVelocityFigGeneral}). Then as $r$ decreases from the minimum of the potential, $\left\vert \dot{r}_{-}\right\vert$ starts decreasing because $V_{m,L}(r)$ is positive and a decreasing function of $r$ between $V_{m,L}(r_{\max}(L))$ and $V_{m,L}(r_{\min}(L))$. Before reaching the peak of the potential
barrier, $\dot{r}_-$ becomes zero at the turning point $r = r_{scat}(L)$ where $E^{2}=V_{m,L}(r_{scat}(L)
)$. At this point, the curve connects smoothly to the curve $\dot{r}_+$ and $\dot{r}_+$ starts growing as $r$\ increases (see the dashed blue curve in Fig. \ref{radialVelocityFigGeneral}).
\end{enumerate}

The dotted red curve corresponds to the separatrix of the two cases (I)-(II) and illustrates the situation for $L =L
_{c}( E) $\ with $E^{2}=V_{m,L_c}(r(L _c) )$. Therefore, we see that the second case describes the trajectories of the particles coming from infinity, but in contrast to the first case, they carry enough angular momentum to be reflected at the potential barrier and escape to infinity. In both cases the topology of these curves is $\mathbb{R}$.\\

{\it Step 4:} In the final step of the proof of proposition~\ref{Prop:GammaProp} we show that the one-forms $dF_{0}$, $dF_{1}$, $dF_{2}$, and $dF_{3}$ are linearly independent at each point of the invariant set $\Gamma_{m,E,L _{z},L}$, provided $0 < |L_z| < L$ and $\{ E,L \}$ does not correspond to a circular orbit. In particular, if $E > m$, then all circular orbits correspond to maxima of the effective potential and it is sufficient to require that $0 < |L_z| < L$ and $L\neq L_c(E)$.

Since the differentials $dF_{0}$, $dF_{1}$, $dF_{2}$, and $dF_{3}$ are the co-normals to the invariant sets $\Gamma _{m,E,L
_{z},L }$, their linear independency implies that the latter are smooth $4$-dimensional submanifold of $\Gamma$ with the
corresponding tangent vector fields $X_{F_{0}}$, $X_{F_{1}}$, $X_{F_{2}}$,
and $X_{F_{3}}$ that are also linearly independent. Just as in the proof of Proposition~2 in~\cite{pRoS16}, it follows from this that the restriction of the Poincar\'e one-form to $\Gamma_{m,E,L _{z},L}$ is closed, and this concludes the proof of Proposition~\ref{Prop:GammaProp}.

\begin{lemma}
\label{Lem:Differentials} Let $E,m>0$ and $L_z\neq 0$. The differentials $dF_{0}$, $dF_{1}$, 
$dF_{2}$, and $dF_{3}$ are linearly independent from each other unless one
of the following two cases occur: 

\begin{enumerate}
\item[(a)] Motion confined to the equatorial plane: $|L_z| = L$,

\item[(b)] Circular trajectories: 
\begin{equation}
V_{m,L }(r)=E^{2},\qquad \frac{d}{dr}V_{m,L }(r)=0.
\end{equation}
\end{enumerate}
\end{lemma}

\proofof{Lemma~\ref{Lem:Differentials}} First, we note that the differential forms $dF_{1}=-dp_{t}\neq 0$ and $dF_{2}=dp_{\varphi }\neq 0$ are linearly independent from each other. Next, we calculate
\begin{equation}
dF_{3}=2\left( p_{\vartheta }dp_{\vartheta }-\frac{L _{z}^{2}}{\sin
^{2}\vartheta }\cot \vartheta d\vartheta +L _{z}\frac{dF_{2}}{\sin
^{2}\vartheta }\right),
\end{equation}
which shows that $dF_{2}$ and $dF_{3}$ are linearly independent from each other
unless $p_{\vartheta }=0$ and $\vartheta =\pi/2$, which is characteristic of the motion in the equatorial plane given in case $(a)$. Next, by expressing the last differential form $dF_{0}$ as follows:
\begin{eqnarray}
dF_{0} &=&\left[ \frac{2-\sigma \left( r\right) \psi (r)}{\sigma \left(
r\right) }E+\frac{1-\sigma \left( r\right) \psi (r)}{\sigma \left( r\right) }%
p_{r}\right] dF_{1} 
 + \left[ \frac{1-\sigma \left( r\right) \psi (r)}{\sigma
\left( r\right) }E-\psi (r)p_{r}\right] dp_{r}  \notag \\
&-& \left[ \frac{\psi ^{\prime }(r)}{2}(E+p_{r})^{2}+\frac{\sigma ^{\prime
}\left( r\right) }{\sigma ^{2}\left( r\right) }(E+p_{r})E-\frac{L ^{2}}{%
r^{3}}\right] dr-\frac{1}{2r^{3}}dF_{3},
\end{eqnarray}%
we find that $dF_{0}$ is linearly independent from $dF_{1}$, $dF_{2}$, and $dF_{3}$ unless 
\begin{equation}
\frac{1-\sigma \left( r\right) \psi (r)}{\sigma \left( r\right) }E-\psi
(r)p_{r}=0,  \label{ZeroRadialVelocityCondition}
\end{equation}
and
\begin{equation}
\frac{\psi ^{\prime }(r)}{2}(E+p_{r})^{2}+\frac{\sigma ^{\prime }\left(
r\right) }{\sigma ^{2}\left( r\right) }(E+p_{r})E-\frac{L ^{2}}{r^{3}}=0.
\label{StableCircularOrbit}
\end{equation}
According to Eq.~(\ref{Eq:rPlaneRestrictionBis}), the first relation is equivalent to the condition $V_{m,L }(r)=E^{2}$. Next, by
introducing Eqs.~(\ref{Eq:rPlaneRestriction}) and (\ref%
{ZeroRadialVelocityCondition}) into the expression for the derivative of the effective potential (Eq.~(\ref{DofV})), one obtains an equation which is equivalent to the second relation~(\ref{StableCircularOrbit}). Therefore, the conditions~(\ref{ZeroRadialVelocityCondition})-(\ref{StableCircularOrbit}) are
equivalent to $(b)$. \hfill \fbox{} \vspace{0.3cm}

\textbf{Remarks}:

\begin{enumerate}
\item The results obtained in this subsection show that the phase space $\Gamma$ naturally splits into four invariant components:
\begin{equation}
\Gamma = \Gamma_{abs} \cup 
\Gamma_{ej} \cup
\Gamma_{scat} \cup \Gamma_{0}.
\end{equation}
Here, $\Gamma_{abs}$ corresponds to absorbed particles emanating from infinity, that is, it consists of the union of those invariant manifolds $\Gamma _{m,E,L _{z},L}$ which have constants of motion with values $E > m > 0$, $|L_z| < L$ and $L < L_c(E)$, and for which the particles are infalling ($p_r = p_{r-}$ in Eq.~(\ref{Eq:prpm})). Similarly, $\Gamma_{ej}$ corresponds to particles that are ejected from the white hole and reach the asymptotic region, which are characterized by the same bounds as in the previous case, but for which the particles are outgoing ($p_r = p_{r+}$ in Eq.~(\ref{Eq:prpm})). Next, $\Gamma_{scat}$ corresponds to scattered particles emanating from infinity, for which $E > m > 0$, $|L_z| < L$ and $L > L_c(E)$ (in this case both signs in $p_r = p_{r\pm}$ are possible). $\Gamma_0$ contains all other types of orbits, including bound orbits, orbits emanating from the white hole which are scattered at the effective potential and fall into the black hole, and the special orbits described in Lemma~\ref{Lem:Differentials}.

\item In the rest of this study, we shall focus on the submanifold $\Gamma _{accr} := \Gamma_{abs}\cup\Gamma_{scat}\subset \Gamma$ consisting of absorbed and scattered trajectories, which are the relevant components to describe the accretion of a collisionless kinetic
gas into the general spherically symmetric and static black hole
spacetimes.
\end{enumerate}

\subsection{New symplectic coordinates and the most general solution of the
collisionless Boltzmann equation}

In this section, we briefly discuss how the most general solution of the collisionless Boltzmann equation can be found in the case of general
static and spherically symmetric black hole spacetimes satisfying the conditions (i)--(iv). The methodology follows closely the one in Ref.~\cite{pRoS16} for the standard Schwarzschild case, and hence we refer the reader to that reference for more details.

Following~\cite{pRoS16}, one can introduce new symplectic local coordinates $(Q^{\mu },P_{\mu })$ on the invariant submanifold $\Gamma _{accr}$ from the generating function~\cite{Arnold-Book}
\begin{equation}
S(x;m,E,L _{z},L ):=\int\limits_{\gamma _{x}}\Theta ,
\label{Eq:GeneratingFunction}
\end{equation}
which is defined as the line integral of the Poincar\'{e} one-form $\Theta $ over a curve $\gamma _{x}$ within the invariant submanifold $\Gamma_{m,E,L _{z},L }$ which connects a fixed reference point to a point over the fiber over $x$.
Note that deformations of the curve $\gamma _{x}$ within $\Gamma_{m,E,L _{z},L }$ do not affect the value
of $S(x;m,E,L _{z},L )$, because the restriction of the Poincar\'{e} one-form to $\Gamma_{m,E,L _{z},L }$ is closed, as follows from Proposition~\ref{Prop:GammaProp}.

The generating function $S$ yields new symplectic coordinates $(Q^{\mu
},P_{\mu })$ where, by definition, the $P_{\mu }$ coordinates are given by the integrals of motion, that is,
\begin{eqnarray}
P_{0}(x,p):= &&\sqrt{2F_{0}}=\sqrt{-2H}=m,  \label{P_3} \\
P_{1}(x,p):= &&F_{1}=-p_{t}=E, \\
P_{2}(x,p):= &&F_{2}=p_{\varphi }=L _{z}, \\
P_{3}(x,p):= &&\sqrt{F_{3}}=\sqrt{p_{\vartheta }^{2}+\frac{p_{\varphi }^{2}}{%
\sin ^{2}\vartheta }}=L ,
\end{eqnarray}%
and the $Q^\mu$ coordinates are obtained from
\begin{equation}
Q^{\mu }=\frac{\partial S}{\partial P_{\mu }}.
\label{Q3}
\end{equation}
By construction, it follows that $dp_{\mu }\wedge dx^{\mu }=dP_{\mu}\wedge dQ^{\mu}$ which shows that the coordinates $(Q^{\mu },P_{\mu })$ are new symplectic
coordinates on $\Gamma _{accr}$ in addition to the primary symplectic
coordinates $(x^{\mu },p_{\mu })$. By definition, the coordinates $P_\mu$ parametrize the four-dimensional invariant submanifolds $\Gamma_{m,E,L _{z},L}$ which have topology $\Real^2\times S^1\times S^1$, whereas $Q^\mu$ provide coordinates on each of these invariant submanifolds, with $Q^2$ and $Q^3$ being $2\pi$-periodic angle variables which parametrize the two $S^1$ factors. Furthermore, one can show~\cite{pRoS16} that the vector fields
\begin{equation}
-\frac{\partial}{\partial Q^1},\qquad
\frac{\partial}{\partial Q^2}
\end{equation}
correspond to the complete lifts of the asymptotically timelike and azimuthal Killing vector fields $\frac{\partial}{\partial t}$ and $\frac{\partial}{\partial\varphi}$, respectively. Also, it can be shown that the generators of the rotation group are linear combinations of
\begin{equation}
\frac{\partial}{\partial Q^2},\qquad \frac{\partial}{\partial Q^3},\qquad
\frac{\partial}{\partial P_2}.
\end{equation}
The one-particle Hamiltonian in these
coordinates is given by $H=-P_{0}^{2}/2$, and the corresponding
Hamiltonian vector field acquires the simple form
\begin{equation}
X_{H}=\frac{\partial H}{\partial P_{\mu }}\frac{\partial }{\partial Q^{\mu }}%
-\frac{\partial H}{\partial Q^{\mu }}\frac{\partial }{\partial P_{\mu }}%
=-P_{0}\frac{\partial }{\partial Q^{0}}.
\end{equation}%
Hence, in the new symplectic local coordinates, the collisionless Boltzmann equation takes
the form 
\begin{equation}
\frac{\partial f}{\partial Q^{0}}=0.
\end{equation}
The most general solution for which $f$ is supported on $\Gamma_{accr}$ is given by the closed-form expression
\begin{equation}
f(x,p)=\mathcal{F}(Q^{1},Q^{2},Q^{3},P_{0},P_{1},P_{2},P_{3}),
\label{Eq:SchwCollisionlessf}
\end{equation}
where $\mathcal{F}$ is an arbitrary function of its argument which is $2\pi$-periodic in the angle variables $Q^2$ and $Q^3$. Hence, Eq.~(\ref{Eq:SchwCollisionlessf}) provides the most general collisionless distribution function describing the
accretion process. We summarize the results of this subsection in:

\begin{theorem}[cf. Theorem~1 in~\cite{pRoS16}]
\label{Thm:Main} Consider a generic static and spherically symmetric black hole spacetime of the form (\ref{metric}) satisfying the conditions (i)--(iv) in section~\ref{SubSec:GenericBH}. On the invariant submanifold $\Gamma
_{accr}\subset \Gamma $ of phase space, the most general collisionless
distribution function is given by equation~(\ref{Eq:SchwCollisionlessf}),
where the function $\mathcal{F}$ is $2\pi $-periodic in the angle variables $Q^{2}$ and $Q^{3}$, and the action-angle-like variables $(Q^{\mu },P_{\mu })$
are defined in equations~(\ref{P_3}--\ref{Q3}).

Moreover, the distribution function is stationary if and only if $\mathcal{F}$
is independent of $Q^{1}$, axisymmetric if and only if $\mathcal{F}$ is
independent of $Q^{2}$, and spherically symmetric if and only if $\mathcal{F}
$ is independent of $Q^{2}$, $Q^{3}$, and $P_{2}$.
\end{theorem}

\textbf{Remarks}:
\begin{enumerate}
\item One important consequence of  this theorem is that a stationary and spherically symmetric collisionless kinetic gas on $\Gamma_{accr}$ is described by a distribution function $f$ depending on $(x,p)$ only through the integrals of motion associated with the conserved quantities $m$, $E$, and $L$. The relation of this result with Jeans' theorem is explained in~\cite{pRoS16}.

\item However, generally, a stationary and collisionless distribution function
depends on more variables than the conserved quantities. For instance, a stationary and axisymmetric distribution function can depend on $Q^{3}$ in addition to the conserved quantities $\left( m,E,L ,L _{z}\right) $. This observation has been exploited in Refs.~\cite{pMoA21a,pMoA21b} to construct the kinetic analogue of the Bondi-Hoyle accretion in the relativistic regime.

\item In a similar fashion, symplectic coordinates which trivialize the Hamiltonian vector field on the components $\Gamma_{ej}$ and $\Gamma_0$ could be constructed to obtain an explicit representation of the solution of the collisionless Boltzmann equation on these components. However, this will not be needed for the problems addressed in the present article.
\end{enumerate}


\section{Particle current density and energy-momentum-stress tensor}
\label{Sec:Observables} 

Now, we focus on the relevant observables, namely the particle
current density and the energy-momentum-stress tensor, associated with a
relativistic collisionless kinetic gas on a general static and spherically
symmetric spacetime $(\mathcal{M},g)$ described by the metric (\ref{metric}). As stated in the introduction, we focus on the case in which the individual gas particles have identical mass $m > 0$, such that the one-particle distribution function $f$ leaves on the future mass shell\footnote{In contrast, in the work~\cite{pRoS16}  the more general situation in which the one-particle distribution function was defined on the whole cotangent bundle was considered. In that work, the distribution function $f$ had units $(length)^{-8}$ instead of $(length)^{-6}$.} and has units $(length)^{-6}$. In terms of adapted local coordinates $(x^\mu,p_\mu)$, the observables are defined by
\begin{equation}
J_{\mu }(x) :=\int\limits_{P_x^+(m)} p_{\mu }f(x,p)\mbox{dvol}%
_{x}(p),\qquad T_{\mu \nu }(x):=\int\limits_{P_x^+(m)}p_{\mu }p_{\nu
}f(x,p)\mbox{dvol}_{x}(p),
\end{equation}
where $P_x^+(m)$ refers to the future mass hyperboloid at $x$ and $\mbox{dvol}_{x}(p)$ is the Lorentz-invariant volume element on $P_x^+(m)$, given by (see for instance Appendix~C of Ref.~\cite{rAcGoS22})
\begin{equation}
\mbox{dvol}_{x}(p) := \frac{\sqrt{-\det (g^{\mu \nu }(x))}dp_r\wedge dp_\vartheta\wedge dp_\varphi}{p^t}.
\end{equation}
It can be shown that $J_\mu$ and $T_{\mu\nu}$ are divergence-free provided $f$ satisfies the collisionless Boltzmann equation. Furthermore, as long as $f > 0$, $J^\mu$ is future-directed timelike and $T_{\mu\nu}$ satisfies all the standard energy conditions (dominant, strong, weak, null). For a proof of these statements, see for example Refs.~\cite{jE71,oStZ13,rAcGoS22}.

Through the next subsections, we shall present these relations more
explicitly in terms of the new symplectic coordinates on $\Gamma_{accr}$ introduced in the previous section and in terms of the coefficients parametrizing the generic metric. We will employ these new expressions to investigate the accretion problem in the next section.

\subsection{Explicit expressions for the observables}

In order to calculate the observables, first, we choose the reference point in Eq.~(\ref{Eq:GeneratingFunction}) to lie on the horizon $(r_{h},p_{r-}(r_{h}))$ in case (I), whereas we consider the reference point to be the turning point $(r_{scat},p_{r+}(r_{scat})=p
_{r-}(r_{scat}))$\ satisfying $V_{L}(r_{scat})=E ^{2}$ in case
(II). In order to compute the observables, we find it convenient to
introduce an angle $\varpi$ such that 
\begin{eqnarray}
p _{\vartheta } &=&L \cos \varpi , \\
\frac{L_{z}}{\sin \vartheta } &=&L \sin \varpi .
\end{eqnarray}%
The volume element in terms of the new quantities reads
\begin{equation}
\mbox{dvol}_{x}(p) = \frac{dE (L dL )d\varpi}{
r^{2}\sqrt{E ^{2}-V_{L}(r)}}.  \label{Eq:FibreMeasure}
\end{equation}

By considering the cases of (I) absorbed particles and (II)
scattered particles discussed in the previous section, we split the fiber
integral into two parts corresponding to these types of trajectories. Accordingly, the observables are written as a sum over these two parts. Whereas the range for $\varpi$ is $\varpi \in (0,2\pi )$, the ranges for the energy $E$ and angular momentum $L$ are as follows:

\begin{enumerate}
\item[(I)] In the case of absorbed particles, $E > m$ and 
$0 < L < L _{c}(E)$. Here, $L _{c}(E)$ is the critical angular
momentum defined in Eq.~(\ref{lCritCond}) and the particles fall into the black hole for $L <L _{c}(E)$. Therefore, in the fiber integrals, we should only consider the solution $p_{r-}(r)$ from
the second expression in Eq.~(\ref{Eq:prpm}) that is regular at the event horizon $r=r_{h}$.

\item[(II)] In the case of scattered particles, since $E$ and $L$
are limited by multiple restrictions such as the position $r$ of the fiber,
the ranges of energy and angular momentum are more complicated compared to
the previous case. Firstly, the lower bound $L >L _{c}(E)$ on the
angular momentum is required for the particles to be reflected at the
centrifugal barrier. Secondly, in this case, the particles are coming from
infinity and are reflected before reaching the peak of the potential.
Therefore, the scattering point $r=r_{scat}$ has to be located at $r_{scat}>r_{\max }(L )$, beyond the position of the potential's maximum. Finally, $r$ should follow
the condition $V_{m,L }(r)\leq E^{2}$ in order to be in the allowed
region.

The explicit limits on $E$ and $L $ which fulfill these restrictions are given in the next lemma. Note that for computing the fiber integrals in this case, we have to sum over both solutions $p_{r+}(r)$ and $p_{r-}(r)$ presented in equation~(\ref{Eq:prpm}).
\end{enumerate}

\begin{lemma}
\label{Lem:RangeEmin} 
Let $r>r_{h}$ and $E>m$, and define 
\begin{equation}
E_{\min }(r):=\left\{ 
\begin{array}{rl}
\infty , & r\leq r_{ph}, \\ 
\sqrt{\chi \left( r\right) \left( m^{2}+\frac{\tilde{L}^{2}\left(
r\right) }{r^{2}}\right) }, & r_{ph}<r<r_{mb}, \\ 
m, & r\geq r_{mb}.
\end{array}
\right. \qquad L _{\max }(E,r):=r\sqrt{\frac{E^{2}}{\chi \left( r\right) }%
-m^{2}},  \label{Eq:EminLambdaMaxDef}
\end{equation}
where $\tilde{L}(r)$ is the value for the total angular momentum for which the effective potential has its maximum at $r$, that is, $r_{max}(\tilde{L}(r)) = r$ (see Eq. (\ref{GeneralRminRmax})), and $r_{mb}$ is the inner radius of the marginally bound orbit whose energy is $E = m$.

Then, the ranges corresponding to case (II) are: 
\begin{equation}
E>E_{\min }(r),\qquad L _{c}(E)<L <L _{\max }(E,r).
\end{equation}
\end{lemma}

\textbf{Remark}: By differentiating $V_{m,L}$ one obtains an explicit formula for $\tilde{L}$:
\begin{equation}
\tilde{L}^2(r) = \frac{m^2 r^3\chi^\prime(r)}{2\chi(r)-r\chi^\prime(r)},
\label{Eq:Ltilde}
\end{equation}
which yields
\begin{equation}
E_{min}(r) = \frac{m\chi(r)}{\sqrt{\chi(r) - \frac{1}{2} r\chi'(r)}},\quad
r_{ph}<r<r_{mb}.
\end{equation}
For instance, in the Schwarzschild case, this reduces to
\begin{equation}
E_{min}(r) = m\left( 1 - \frac{2M}{r}\right)\sqrt{\frac{r}{r-3M}},\qquad
r_{ph} = 3M < r < 4M = r_{mb},
\end{equation}
which agrees with the corresponding expression in Eq.~(50) of~\cite{pRoS16}.\\

\proofof{Lemma~\ref{Lem:RangeEmin}}
We first note that $V_{L _{\max }}(r)=E^{2}$,
thus the restriction $L <L _{\max }$ leads to the condition $V_{L
}(r)<E^{2}$. Therefore, the inequalities $L _{c}(E)<L <L _{\max
}(E,r)$ are necessary for case (II). Note that we can explicitly calculate $L _{\max }(E,r)$ from the relation $V_{L _{\max }}(r)=E^{2}
$ which yields the expression found in Eq.~(\ref{Eq:EminLambdaMaxDef}).

To justify the first condition $E>E_{\min }(r)$, we note that when $L $
increases from $L _{ISCO}$ to $\infty $, $r_{\max }(L )$ decreases
monotonically from $r_{ISCO}$ to $r_{ph}$ (see Fig.~\ref{GfunctionFig} and
related discussion in the proof of Lemma \ref{Lem:Potential}). Therefore, the condition $r>r_{\max }(L )$ is never
satisfied for $r\leq r_{ph}$ and the integration region is empty, so in this case we set $E_{min}(r) = \infty$.

Next, we claim that for $r > r_{mb}$, the condition $r>r_{\max}(L)$ is automatically satisfied. This follows from the already established monotonicity properties of $r_{max}(L)$ and $L_c(E)$ which imply that $r_{max}(L)\leq r_{max}(L_c(m))$ since $L > L_c(E) > L_c(m)$. On the other hand, $r_{mb}$ is characterized by the condition $r_{mb}=r_{\max }(L _{c}(m))$. Hence, it follows that $r_{max}(L)\leq r_{mb}$.

Finally, for $r_{ph}<r<r_{mb}$, the condition $r>r_{\max }(L)$ is equivalent to a
conditional inequality $L > \tilde{L}(r)$ for the angular momentum, and in turn implies  
\begin{equation}
E^{2}\geq V_{L }(r)>\chi \left( r\right) \left( m^{2}+\frac{\tilde{L}%
^{2}\left( r\right) }{r^{2}}\right),
\end{equation}%
where the right-hand side is equal to $E_{\min }^{2}(r)$.
\qed \\

As the next step, we rescale the particle's mass $m$ in order to simplify the computation of the fiber integrals, that is, we write
\begin{equation}
p_{r\pm }=m\pi _{r\pm },\quad p_{\vartheta }=m\pi _{\vartheta },\quad
E=m\varepsilon ,\quad L =m\lambda ,\quad L _{z}=m\lambda _{z},
\end{equation}
with the radial momentum 
\begin{equation}
\pi _{r\pm }(r):=\frac{\left[ 1-\sigma \left( r\right) \psi (r)\right]
\varepsilon \pm \sqrt{\varepsilon ^{2}-U_{\lambda }(r)}}{\sigma \left(
r\right) \psi (r)}=\frac{\sigma \left( r\right) \left( 1+\frac{\lambda ^{2}}{%
r^{2}}\right) -\left[ 2-\sigma \left( r\right) \psi (r)\right] \varepsilon
^{2}}{\left[ 1-\sigma \left( r\right) \psi (r)\right] \varepsilon \mp \sqrt{%
\varepsilon ^{2}-U_{\lambda }(r)}},  \label{Eq:pxipm}
\end{equation}
and the rescaled effective potential
\begin{equation}
U_{\lambda }(r):=\frac{V_{m,L }(r)}{m^{2}}=\chi (r)\left( 1+\frac{\lambda
^{2}}{r^{2}}\right),
\label{Eq:Ulambda}
\end{equation}
which leads to the following expression for the volume element (\ref{Eq:FibreMeasure}):
\begin{equation}
\mbox{dvol}_{x}(p) = \frac{m^2 d\varepsilon (\lambda d\lambda)d\varpi}{
r^{2}\sqrt{\varepsilon ^{2}-U_{\lambda}(r)}}. \label{Eq:FibreMeasureRescaled}
\end{equation}

Note that only the quantities $\pi _{r_{\pm }}$, $\varepsilon$, and $U_{\lambda }(r)$ are dimensionless; in contrast $\pi_\vartheta$, $\lambda$, and $\lambda_z$ have units of mass.\footnote{Note that the quantities $\pi_\vartheta$, $\lambda$, and $\lambda_z$ differ from the corresponding quantities defined in Ref. \cite{pRoS16} by the black hole mass factor.}

After these remarks, we are ready to calculate the observable
quantities. To this purpose, we first note that the fiber integral 
\begin{equation}
I(x):=\int\limits_{P_x ^{+}(m)}f(x,p)\mbox{dvol}_{x}(p),
\end{equation}%
consists of the following two parts 
\begin{equation}
I(x)=I^{(abs)}(x)+I^{(scat)}(x),
\end{equation}%
that take into account the contribution from the absorbed particles 
\begin{equation}
I^{(abs)}(x)=\frac{m^2}{r^{2}}\int\limits_{1}^{\infty }\int\limits_{0}^{2\pi
}\int\limits_{0}^{\lambda _{c}(\varepsilon )}\left. \mathcal{I}(Q^{1},Q^{2},Q^{3},\varepsilon ,\lambda_z 
,\lambda )\right\vert _{\pi _{r}=\pi _{r-}(r)}\frac{\lambda d\lambda d\varpi
d\varepsilon }{\sqrt{\varepsilon ^{2}-U_{\lambda }(r)}},
\label{Eq:JgeneralinQ's}
\end{equation}
and the contribution from the scattered particles
\begin{equation}
I^{(scat)}(x) =\frac{m^2}{r^{2}}\int\limits_{\varepsilon _{min}(r)}^{\infty
}\int\limits_{0}^{2\pi }\int\limits_{\lambda _{c}(\varepsilon )}^{\lambda
_{max}(\varepsilon ,r)}\left. 
\sum\limits_{\pm}\mathcal{I}(Q^{1},Q^{2},Q^{3},\varepsilon
,\lambda _z ,\lambda )\right\vert _{\pi _{r}=\pi
_{r\pm}(r)}\frac{\lambda d\lambda d\varpi d\varepsilon }{\sqrt{\varepsilon
^{2}-U_{\lambda }(r)}},
\label{Eq:JgeneralscatQ's}
\end{equation}
where $\sum_{\pm }$ stands for the sum over the contributions from $\pi _{r-}$\ and $\pi _{r+}$. Here, the variables $Q^\mu$ can be computed by means of Eq.~(\ref{Q3}), $\varepsilon _{min}(r) = E_{min}(r)/m$, $\lambda_c(\varepsilon) = L_c(E)/m$, and $\lambda_{max}(\varepsilon,r) = L_{max}(E,r)/m$, and the function $\mathcal{I}$ is related to the function $\mathcal{F}$ in Theorem~\ref{Thm:Main} according to $\mathcal{I}(Q^1,Q^2,Q^3,\varepsilon
,\lambda _z ,\lambda) = \mathcal{F}(Q^1,Q^2,Q^3,m,E,L_z,L)$. By considering the relations $p_t=-m\varepsilon$, $p_{r}=m\pi _{r}$, $p_{\vartheta
}=m\lambda \cos \varpi $, and $p_{\varphi }=m\lambda \sin \varpi \sin
\vartheta $, we can compute the general expressions for the current density
and energy-momentum-stress tensor in a similar fashion.

\subsection{Observables in the spherically symmetric case}

For a spherically symmetric distribution function, $\mathcal{I}$ is independent of $Q^{2}$, $Q^{3}$, and $\lambda_{z}$ (See Theorem~\ref{Thm:Main}), hence the observable quantities associated with the absorbed
particles simplify to 
\begin{eqnarray}
J_{a}^{(abs)}(t,r) &=&\frac{2\pi m^3}{r^{2}}\int\limits_{1}^{\infty
}\int\limits_{0}^{\lambda _{c}(\varepsilon )}u_{a-}\left. \mathcal{I}(Q^{1},\varepsilon ,\lambda )\right\vert _{\pi _{r}=\pi _{r-}(r)}\frac{%
\lambda d\lambda d\varepsilon }{\sqrt{\varepsilon ^{2}-U_{\lambda }(r)}}%
,\qquad a=t,r,  \label{Eq:Jain} \\
T_{ab}^{(abs)}(t,r) &=&\frac{2\pi m^4}{r^{2}}\int\limits_{1}^{\infty
}\int\limits_{0}^{\lambda _{c}(\varepsilon )}u_{a-}u_{b-}\left. \mathcal{I}(Q^{1},\varepsilon ,\lambda )\right\vert _{\pi _{r}=\pi _{r-}(r)}\frac{%
\lambda d\lambda d\varepsilon }{\sqrt{\varepsilon ^{2}-U_{\lambda }(r)}}%
,\qquad a,b=t,r,  \label{Eq:Tabin} \\
T_{\vartheta \vartheta }^{(abs)}(t,r) &=&\frac{\pi m^4}{r^{2}}%
\int\limits_{1}^{\infty }\int\limits_{0}^{\lambda _{c}(\varepsilon )}\lambda
^{2}\left. \mathcal{I}(Q^{1},\varepsilon ,\lambda )\right\vert _{\pi
_{r}=\pi _{r-}(r)}\frac{\lambda d\lambda d\varepsilon }{\sqrt{\varepsilon
^{2}-U_{\lambda }(r)}},  \label{Eq:Tthetathetain}
\end{eqnarray}%
where $(u_{t\pm },u_{r\pm })=(-\varepsilon ,\pi _{r\pm }(r))$, $T_{\varphi
\varphi }^{(abs)}=\sin ^{2}\vartheta \,T_{\vartheta \vartheta }^{(abs)}$,
and the remaining components of $J_{\mu }^{(abs)}$ and $T_{\mu \nu }^{(abs)}$
are zero. The nonvanishing components of the scattering part are given by 
\begin{eqnarray}
J_{a}^{(scat)}(t,r) &=&\frac{2\pi m^3}{r^{2}}\int\limits_{\varepsilon
_{min}(r)}^{\infty }\int\limits_{\lambda _{c}(\varepsilon )}^{\lambda
_{max}(\varepsilon ,r)}\sum\limits_{\pm }\left[ u_{a\pm }\left. \mathcal{I}(Q^{1},\varepsilon ,\lambda )\right\vert _{\pi _{r}=\pi _{r\pm }(r)}%
\right] \frac{\lambda d\lambda d\varepsilon }{\sqrt{\varepsilon
^{2}-U_{\lambda }(r)}},  \label{Eq:Jascat} \\
T_{ab}^{(scat)}(t,r) &=&\frac{2\pi m^4}{r^{2}}\int\limits_{\varepsilon
_{min}(r)}^{\infty }\int\limits_{\lambda _{c}(\varepsilon )}^{\lambda
_{max}(\varepsilon ,r)}\sum\limits_{\pm }\left[ u_{a\pm }u_{b\pm }\left. 
\mathcal{I}(Q^{1},\varepsilon ,\lambda )\right\vert _{\pi _{r}=\pi
_{r\pm }(r)}\right] \frac{\lambda d\lambda d\varepsilon }{\sqrt{\varepsilon
^{2}-U_{\lambda }(r)}},  \label{Eq:Tabscat} \\
T_{\vartheta \vartheta }^{(scat)}(t,r) &=&\frac{\pi m^4}{r^{2}}%
\int\limits_{\varepsilon _{min}(r)}^{\infty }\int\limits_{\lambda
_{c}(\varepsilon )}^{\lambda _{max}(\varepsilon ,r)}\lambda
^{2}\sum\limits_{\pm }\left[ \left. \mathcal{I}(Q^{1},\varepsilon
,\lambda )\right\vert _{\pi _{r}=\pi _{r\pm }(r)}\right] \frac{\lambda
d\lambda d\varepsilon }{\sqrt{\varepsilon ^{2}-U_{\lambda }(r)}},
\label{Eq:Tthetathetascat}
\end{eqnarray}%
where $T_{\varphi \varphi }^{(scat)}=\sin ^{2}\vartheta \,T_{\vartheta
\vartheta }^{(scat)}$. Here, the variable $Q^1$ obtained from Eq.~(\ref{Q3}) is given by
\begin{equation}
\left. Q^1 \right|_{\pi_r = \pi_{r\pm}} = -\bar{t} \pm E\int^r\frac{1}{\sqrt{E^2 - V_{m,L}(r)}}\frac{dr}{\sigma(r)\psi(r)},
\end{equation}
where $\bar{t}$ is the Schwarzschild time which is related to the time
coordinate $t$ via the transformation~(\ref{Eq:ttbar}). Note that $Q^1\to -\infty$ as $t\to \infty$, and hence the observables are expected to converge in time to those obtained by replacing $\mathcal{I}(Q^{1},\varepsilon,\lambda )$ with its limit for $Q^1\to -\infty$. Using Lebesgue's dominated convergence theorem, this expectation can be shown to be fulfilled provided suitable assumptions on the function $\mathcal{I}$ are made (see~\cite{pRoS16} for details). For this reason, in the following we shall further specialize to the case for which $\mathcal{I}$ is independent of $Q^1$, describing an equilibrium (steady-state) situation according to Theorem~\ref{Thm:Main}.

\subsection{Example of a steady-state, spherically symmetric isotropic gas}
\label{SubSec:ExSSSI}

As an explicit example, consider a spherically symmetric collisionless gas which gives rise to a steady-state configuration which is isotropic in the asymptotic region. In this case, $\mathcal{I}$ is also independent of $\lambda$. Therefore, the one-particle distribution function reduces to
\begin{equation}
f(x,p)=\left. f_{\infty }(\varepsilon
)\right\vert _{\varepsilon =P_{1}(x,p)/m},  
\label{Eq:SSSSIG}
\end{equation}
on $\Gamma_{accr}$, with $f_{\infty }:(1,\infty )\rightarrow \mathbb{R}$ being a smooth and
non-negative function that is bounded and decays sufficiently fast at
infinity. Since $f(x,p)$ only depends on the energy variable $\varepsilon$, we can explicitly integrate the relations~(\ref{Eq:Jain}--\ref{Eq:Tthetathetascat}) over the total angular momentum $\lambda $, reducing the fiber integrals to single integrals. By employing the integral identities presented in App.~\ref{App:integrals}, one obtains 
\begin{eqnarray}
\left( 
\begin{array}{l}
J^{t} \\ 
J^{r}%
\end{array}%
\right) ^{(abs)}(r) &=&\frac{2\pi m^{3}}{r^{2}}\int\limits_{1}^{\infty }%
\frac{\lambda _{c}^{2}(\varepsilon )}{\sqrt{\varepsilon ^{2}-U_{0}(r)}+\sqrt{%
\varepsilon ^{2}-U_{c}(r)}}\left( 
\begin{array}{l}
v^{t} \\ 
v^{r}%
\end{array}%
\right) f_{\infty }(\varepsilon )d\varepsilon ,  \label{Eq:JainSS} \\
\left( 
\begin{array}{ll}
T^{t}{}_{t} & T^{t}{}_{r} \\ 
T^{r}{}_{t} & T^{r}{}_{r}%
\end{array}%
\right) ^{(abs)}(r) &=&\frac{2\pi m^{4}}{r^{2}}\int\limits_{1}^{\infty }%
\frac{\lambda _{c}^{2}(\varepsilon )}{\sqrt{\varepsilon ^{2}-U_{0}(r)}+\sqrt{%
\varepsilon ^{2}-U_{c}(r)}}\left( 
\begin{array}{ll}
v^{t}v_{t} & v^{t}v_{r} + \left[ \sigma(r) - \chi(r) \right] W \\ 
v^{r}v_{t} & v^{r}v_{r}+\chi(r) W
\end{array}
\right) f_{\infty }(\varepsilon )d\varepsilon ,  \label{Eq:TabinSS} \\
(T^{\vartheta }{}_{\vartheta })^{(abs)}(r) &=&\frac{\pi m^{4}}{3r^{4}}%
\int\limits_{1}^{\infty }\lambda _{c}^{4}(\varepsilon )\frac{2\sqrt{%
\varepsilon ^{2}-U_{0}(r)}+\sqrt{\varepsilon ^{2}-U_{c}(r)}}{\left( \sqrt{%
\varepsilon ^{2}-U_{0}(r)}+\sqrt{\varepsilon ^{2}-U_{c}(r)}\right) ^{2}}%
f_{\infty }(\varepsilon )d\varepsilon ,  \label{Eq:TthetathetainSS}
\end{eqnarray}%
where we have introduced the shorthand notation $U_{c}(r):=U_{\lambda
_{c}(\varepsilon )}(r)$. Moreover, the quantity $W$ and the two-vector $%
\left( v^{t},v^{r}\right) $ with corresponding co-vector $\left(
v_{t},v_{r}\right) $\ have the following explicit forms 
\begin{equation}
W:=\frac{1}{12 r^{4}}
\frac{\lambda _{c}^{4}(\varepsilon )}{\left( \sqrt{\varepsilon ^{2}-U_{0}(r)}%
+\sqrt{\varepsilon ^{2}-U_{c}(r)}\right) ^{2}},
\end{equation}
\begin{equation}
\left( 
\begin{array}{l}
v^{t} \\ 
v^{r}%
\end{array}%
\right) =\left( 
\begin{array}{l}
\frac{\varepsilon }{\sigma (r)}+\frac{1-\sigma (r)\psi (r)}{2}\left( \frac{1}{\varepsilon +\sqrt{\varepsilon ^{2}-U_{0}(r)}}+\frac{1+\frac{\lambda
_{c}^{2}(\varepsilon )}{r^{2}}}{\varepsilon +\sqrt{\varepsilon ^{2}-U_{c}(r)}
}\right)  \\ 
-\frac{1}{2\sigma (r)}\left( \sqrt{\varepsilon ^{2}-U_{0}(r)}+\sqrt{%
\varepsilon ^{2}-U_{c}(r)}\right) 
\end{array}%
\right) ,
\end{equation}%
\begin{equation}
\left( 
\begin{array}{l}
v_{t} \\ 
v_{r}%
\end{array}%
\right) =\left( 
\begin{array}{l}
-\varepsilon  \\ 
-\varepsilon +\frac{\sigma (r)}{2}\frac{1}{\varepsilon +\sqrt{\varepsilon
^{2}-U_{0}(r)}}+\frac{\sigma (r)}{2}\frac{1+\frac{\lambda
_{c}^{2}(\varepsilon )}{r^{2}}}{\varepsilon +\sqrt{\varepsilon ^{2}-U_{c}(r)}
}
\end{array}
\right) .
\end{equation}

For the analogous terms corresponding to the scattered particles, we find 
\begin{eqnarray}
\left( 
\begin{array}{l}
J^{t} \\ 
J^{r}%
\end{array}%
\right) ^{(scat)}(r) &=&\frac{4\pi m^{3}}{\chi ^{2}(r)}\int\limits_{%
\varepsilon _{min}(r)}^{\infty }\left( 
\begin{array}{r}
1 \\ 
0%
\end{array}%
\right) \varepsilon \sqrt{\varepsilon ^{2}-U_{c}(r)}f_{\infty }(\varepsilon
)d\varepsilon ,  \label{Eq:JascatSS} \\
\left( 
\begin{array}{ll}
T^{t}{}_{t} & T^{t}{}_{r} \\ 
T^{r}{}_{t} & T^{r}{}_{r}%
\end{array}%
\right) ^{(scat)}(r) &=&\frac{4\pi m^{4}}{\chi ^{2}(r)}\int\limits_{%
\varepsilon _{min}(r)}^{\infty }\left( 
\begin{array}{ll}
-\varepsilon ^{2} & \frac{1-\sigma (r)\psi (r)}{3}\frac{4\varepsilon
^{2}-U_{c}(r)}{\sigma (r)\psi (r)} \\ 
0 & \frac{1}{3}\left[ \varepsilon ^{2}-U_{c}(r)\right]%
\end{array}%
\right) \sqrt{\varepsilon ^{2}-U_{c}(r)}f_{\infty }(\varepsilon
)d\varepsilon ,  \label{Eq:TabscatSS} \\
(T^{\vartheta }{}_{\vartheta })^{(scat)}(r) &=&\frac{2\pi m^{4}}{\chi ^{2}(r)%
}\int\limits_{\varepsilon _{min}(r)}^{\infty }\left\{ \chi (r)\frac{\lambda
_{c}^{2}(\varepsilon )}{r^{2}}+\frac{2}{3}\left[ \varepsilon ^{2}-U_{c}(r)%
\right] \right\} \sqrt{\varepsilon ^{2}-U_{c}(r)}f_{\infty }(\varepsilon
)d\varepsilon .  \label{Eq:TthetathetascatSS}
\end{eqnarray}

It is worth mentioning that the observables $J_{\mu }^{(abs)}$ and $T_{\mu
\nu }^{(abs)}$ from absorbed particles\ are regular at the event horizon $r_{h}$. In contrast, the contribution of the observables $J_{\mu }^{(scat)}$
and $T_{\mu \nu }^{(scat)}$ from\ scattered particles vanishes for $r<r_{ph}$ inside the photon sphere radius (see the definition of $E_{min}(r)$ in Eq. (\ref{Eq:EminLambdaMaxDef})). Besides, if the function $f_{\infty }$ decays
sufficiently fast as $\varepsilon \rightarrow \infty $, the integrals (\ref{Eq:JainSS}--\ref{Eq:TthetathetascatSS}) are well-defined. In the upcoming section, we will present the physical applications of the results obtained throughout this section.

Finally, as in Ref.~\cite{pRoS16}, it will turn out to be instructive to compare the properties of the observables $J^\mu$ and $T^{\mu\nu}$ to those associated with an isotropic perfect fluid. In the latter, the energy-momentum-stress tensor can be diagonalized as follows:
\begin{equation}
T^\mu{}_\nu = \rho e_0^\mu e_{0\nu} + p_1 e_1^\mu e_{1\nu} + p_2 e_2^\mu
e_{2\nu} + p_3 e_3^\mu e_{3\nu},  \label{Eq:TmunuDecomp}
\end{equation}
where ${\bf e}_0,{\bf e}_1,{\bf e}_2,{\bf e}_3$ refer to an orthonormal basis of vector fields, with $e_0^\mu$ being parallel to $J^\mu$ and the principal pressures being equal to each other: $p_1=p_2=p_3$. Through a concrete example in the next section, we shall show that the 
condition for $e_0^\mu$ to be parallel to $J^\mu$ does not always hold in the kinetic model. Furthermore, we will also exhibit cases in which the principal pressures are not identical to each other.


\section{Application to the accretion problem}

\label{Sec:Accretion} 

In this section, we apply our results to the steady-state spherical accretion of a relativistic and collisionless kinetic gas into a general static and spherically symmetric black hole spacetime. First, we calculate the particle and energy fluxes passing through a sphere of constant radius as well as the mass and energy accretion rates. Then, we explore the asymptotic behavior of the observables and show that the gas behaves as an isotropic perfect fluid at infinity. Next, we evaluate the observable quantities at the event horizon that will be useful for extracting the particle density, energy density, and principal pressures at this location.

As the next stage, we focus our attention to the case for which the distribution function is of the Maxwell-J\"uttner form at infinity, and we provide the most general
steady-state, spherically symmetric solution corresponding to a gas which is in equilibrium at a given temperature at infinity.
Next, we calculate the  accretion rate and compression ratio as well as the particle density, energy density, and
the radial and tangential pressures at the event horizon and spatial infinity. These quantities are obtained for the two special cases of low-temperature limit and ultra-relativistic limit at the horizon. Finally, we shall apply our generic formulae obtained for general static and spherically symmetric backgrounds to two concrete examples of black hole spacetimes, namely the Reissner–Nordstr\"om solutions and quantum-corrected black holes, in order to investigate the effects of their free parameters on the observable quantities, explore their deviations from the standard Schwarzschild black hole, and demonstrate an application of the developed formalism.

\subsection{Accretion rate}

We employ the inverse metric~(\ref{Eq:SchwarzschildInv}) to calculate the particle and energy fluxes passing through a sphere of constant areal radius $R$, and we obtain\footnote{For this calculation we have taken the unit normal to the surface to be outward directed, which results in a negative flux.}
\begin{eqnarray}
j_{n}:= & \left[ 4\pi r^{2}\sigma J^{r} \right]_{r=R} =&-4\pi ^{2} m^{3}
\int\limits_{1}^{\infty }\lambda _{c}^{2}(\varepsilon ) f_\infty(\varepsilon) d\varepsilon ,  \label{j_n} \\
j_{\varepsilon }:= & \left[ -4\pi r^{2}\sigma T^{r}{}_{t} \right]_{r=R} =&-4\pi ^{2}m^{4}\int\limits_{1}^{\infty }\varepsilon \lambda _{c}^{2}(\varepsilon
) f_\infty(\varepsilon) d\varepsilon ,  \label{j_epsilon}
\end{eqnarray}%
where $\lambda _{c}(\varepsilon )$ is the critical angular momentum that can be calculated through equation~(\ref{lCritCond}). It is worthwhile noticing that the particle flux~(\ref{j_n}) and energy flux~(\ref{j_epsilon}) are independent of $R$, which is a consequence of the fact that the solution is time-independent and of the conservation laws 
$\nabla _{\mu }J^{\mu }=0$ and $\nabla _{\mu }(-T^{\mu }{}_{\nu }k^{\nu })=0$. Moreover, the contribution from the incoming particles is completely
balanced by outgoing particles scattered at the potential barrier, and thus only the absorbed particles contribute to $j_{n}$\ and $j_{\varepsilon }$ so
that $(J^{r})^{(scat)}=0=(T^{r}{}_{t})^{(scat)}$ (see Eqs. (\ref{Eq:JascatSS}-\ref{Eq:TabscatSS})). Hence, the mass and energy accretion rates into the central black hole are given by
\begin{eqnarray}
\dot{M} :=& m|j_{n}| 
 =& 4 \pi^2 m^4
\int\limits_{1}^{\infty }\lambda _{c}^{2}(\varepsilon ) f_\infty(\varepsilon) d\varepsilon,
\label{Eq:MassAccretionRate}\\
\dot{\mathcal{E}}:=&|j_{\varepsilon}|=& 4\pi ^{2}m^{4}\int\limits_{1}^{\infty }\varepsilon \lambda _{c}^{2}(\varepsilon
) f_\infty(\varepsilon) d\varepsilon \geq \dot{M}.
\label{Eq:EnergyAccretionRate}
\end{eqnarray}
We will further analyze the properties of $\dot{M}$ and $\dot{\mathcal{E}}$ when considering a Maxwell-J\"uttner form for $f_\infty$ in Section~\ref{SubSec:SSS}.

\subsection{Asymptotic limit ($r\to\infty$)}

Next, we investigate the observable quantities at spatial infinity $r\rightarrow \infty $. We have contributions only from the scattered particles in the asymptotic region, and the gas behaves as an isotropic, relativistic perfect fluid, as we are going to show now. Indeed, by taking the limit $r\rightarrow \infty $ in Eqs.~(\ref{Eq:JascatSS}--\ref%
{Eq:TthetathetascatSS}), recalling the assumptions on the functions $\chi$, $\psi$ and $\sigma$, and considering the facts $U_{c}(r)\rightarrow 1$
and $\varepsilon _{min}(r)\rightarrow 1$,\ we obtain 
\begin{equation}
\lim\limits_{r\rightarrow \infty }J^{\mu }=n_{\infty }u_{\infty }^{\mu
},\qquad \lim\limits_{r\rightarrow \infty }T^{\mu \nu }=\left( \rho_\infty + p_{\infty }\right) u_{\infty }^{\mu }u_{\infty }^{\nu }+p_{\infty
}\eta ^{\mu \nu },
\label{Eq:JTInfty}
\end{equation}
where $\mathbf{u}_{\infty }=\mathbf{k}=\partial_t$ is the four-velocity in
the asymptotic region which is described by the Minkowski metric
with the inverse components $\eta ^{\mu \nu }$. Here, the
particle density $n_{\infty }$, energy density $\rho_\infty$, and
pressure $p_{\infty }$ at infinity can be written as 
\begin{equation}
n_{\infty }(z) = m\mathcal{W}_{1,0}[f_\infty],\qquad
\rho_\infty(z) = m^2\mathcal{W}_{2,0}[f_\infty],\qquad 
p_{\infty }(z)=\frac{1}{3} m^2\mathcal{W}_{0,1}[f_\infty],
\label{Eq:nepsp}
\end{equation}
where for an arbitrary function $\phi$ of $\varepsilon$, the functionals $\mathcal{W}_{n,k}[\phi]$ are defined by (cf. Eq.~(29) in Ref.~\cite{aGaM2023})
\begin{equation}
\mathcal{W}_{n,k}[\phi] := 4\pi m^2\int\limits_1^\infty \phi(\varepsilon)\varepsilon^n(\varepsilon^2-1)^{k+1/2} d\varepsilon 
 = 4\pi m^2\int\limits_{0}^{\infty }\phi(\cosh x)\cosh^n (x)\sinh^{2k+2}(x) dx,\qquad n,k=0,1,2,\ldots
\end{equation}
Using the identity $\mathcal{W}_{0,k+1}[\phi'] = -(2k+3)\mathcal{W}_{1,k}[\phi]$ with $\phi'$ denoting the derivative of $\phi$, one obtains from Eq.~(\ref{Eq:nepsp})
\begin{equation}
\frac{p_\infty}{n_\infty} = -m\frac{\mathcal{W}_{0,1}[f_\infty]}{\mathcal{W}_{0,1}[f_\infty']}.
\label{Eq:povern}
\end{equation}
These expressions will be further specialized later.

\subsection{Properties at the horizon ($r = r_h$)}
\label{SubSec:Horizon}

Recall that the contribution from scattered particles vanishes at the
event horizon, and only $(J^{\mu })^{(abs)}$ and $(T^{\mu }{}_{\nu
})^{(abs)} $ lead to a nonvanishing contribution. The particle density $n_{h}
$, energy density $\rho _{h}$, radial pressure $p_{rad,h}$,\ and tangential
pressure $p_{tan,h}$ are determined by $J^{\mu }=n_{h}u_{h}^{\mu }$ with$\
g_{\mu \nu }u_{h}^{\mu }u_{h}^{\nu }=-1$ as well as the decomposition~(\ref%
{Eq:TmunuDecomp}) which exists for a kinetic gas \cite{Synge2-Book,oStZ13}.
Evaluating the relations given in~(\ref{Eq:JainSS}--\ref{Eq:TthetathetainSS}%
) at the horizon $r_{h}$, and taking into account that $U_{\lambda }(r_{h})=0$ and $\psi(r_h)=0$, we obtain
\begin{equation}
\left. \left( 
\begin{array}{l}
J^{t} \\ 
J^{r}
\end{array}
\right) \right\vert _{r_{h}} 
 = \left( 
\begin{array}{c}
a + b \\ 
-a
\end{array} \right),
\qquad
\left. \left( 
\begin{array}{ll}
T^{t}{}_{t} & T^{t}{}_{r} \\ 
T^{r}{}_{t} & T^{r}{}_{r}
\end{array}
\right) \right\vert _{r_{h}} 
 = \left( 
\begin{array}{cc}
-A - B & -A + H \\ 
A & A-B
\end{array}
\right),
\qquad
\left. T^{\vartheta}{}_{\vartheta} \right\vert _{r_{h}}  = \left. T^{\varphi}{}_{\varphi}\right\vert _{r_{h}} = D,
\label{Eq:ObsHorizon}
\end{equation}
where
\begin{equation}
a = \frac{\pi m^3}{\sigma_h r_h^2}\int\limits_1^\infty \lambda_c^2(\varepsilon) f_\infty(\varepsilon) d\varepsilon,
\qquad
b = \frac{\pi m^3}{r_h^2}\int\limits_1^\infty \frac{\lambda_c^2(\varepsilon)}{2\varepsilon^2}\left( 1 + \frac{\lambda_c^2(\varepsilon)}{2r_h^2} \right) f_\infty(\varepsilon) d\varepsilon,
\label{Defsab}
\end{equation}
and
\begin{eqnarray}
A &=&\frac{\pi m^{4}}{\sigma_h r_{h}^{2}}\int\limits_{1}^{
\infty }\lambda _{c}^{2}(\varepsilon )\varepsilon f_\infty(\varepsilon) d\varepsilon , \label{Aexp} \\
B &=&\frac{\pi m^{4}}{r_{h}^{2}}\int\limits_{1}^{\infty }\frac{%
\lambda _{c}^{2}(\varepsilon )}{2\varepsilon }\left( 1+\frac{\lambda
_{c}^{2}(\varepsilon )}{2r_{h}^{2}}\right) f_\infty(\varepsilon) d\varepsilon ,
\\
H &=&\frac{\pi m^{4}\sigma_h}{ r_{h}^{2}}\int\limits_{1}^{%
\infty }\frac{\lambda_{c}^{2}(\varepsilon )}{4\varepsilon
^{3}}\left[ 1+\left( \frac{\lambda _{c}(\varepsilon )}{r_{h}}\right)^{2} +
\frac{1}{3}\left( \frac{\lambda
_{c}(\varepsilon )}{r_{h}}\right) ^{4}\right] f_\infty(\varepsilon) d\varepsilon
, \\
D &=&\frac{\pi m^{4}}{4r_{h}^{4}}\int\limits_{1}^{\infty }\frac{%
\lambda _{c}^{4}(\varepsilon )}{\varepsilon } f_\infty(\varepsilon) d\varepsilon.
\label{Dexp}
\end{eqnarray}
Then, the observables at the horizon are
\begin{equation}
\left. J^\mu \right\vert _{r_{h}} = n_h u_h^\mu,\qquad
\left. T^{\mu\nu} \right\vert _{r_{h}} = \rho_h e_0^\mu e_0^\nu + p_{rad,h} e_1^\mu e_1^\nu + p_{tan,h}\left( e_2^\mu e_2^\nu
+ e_3^\mu e_3^\nu \right),
\end{equation}
with
\begin{equation}
n_h = \sqrt{2\sigma_h a b},\qquad
u_h = \frac{1}{\sqrt{2\sigma_h a b}}\left[ (a+b)\frac{\partial}{\partial t} - a\frac{\partial}{\partial r} \right],
\label{Eq:ObsHorizon1}
\end{equation}
and
\begin{equation}
\rho_h = B + \sqrt{AH},\qquad
p_{rad,h} = -B + \sqrt{AH},\qquad
p_{tan,h} = D.
\label{Eq:ObsHorizon2}
\end{equation}
Here, the tetrad fields are given by
\begin{eqnarray}
e_0 &=& \frac{1}{\sqrt{2\sigma_h}}\left( \frac{A}{H} \right)^{1/4}\left[ \left( 1 + \sqrt{\frac{H}{A}} \right)\frac{\partial}{\partial t} - \frac{\partial}{\partial r} \right],
\\
e_1 &=& \frac{1}{\sqrt{2\sigma_h}}\left( \frac{A}{H} \right)^{1/4}\left[ \left( 1 - \sqrt{\frac{H}{A}} \right)\frac{\partial}{\partial t} - \frac{\partial}{\partial r} \right],
\\
e_2 &=& \frac{1}{r}\frac{\partial}{\partial\vartheta},\qquad
e_3 = \frac{1}{r\sin\vartheta}\frac{\partial}{\partial\varphi}.
\end{eqnarray}
Note that $n_h$, $\rho_h$, $p_{rad,h}$, and $p_{tan,h}$ are independent of $\sigma_h$. Remarkably, they only depend on the horizon radius $r_h$ and integrals involving the rescaled critical angular momentum $\lambda_c(\varepsilon)$. The latter depends only on the properties of the effective potential which, in turn, only depends on the metric function $\chi(r)$ which determines the norm of the asymptotically timelike Killing vector field $\mathbf{k}$, and thus has a geometric meaning. The integrals defining the quantities $a$, $b$, $A$, $B$, $H$, $D$ are still rather complicated, and for this reason we will further specialize them in what follows.

\subsection{Spherically symmetric,
steady-state configurations with given temperature at infinity}
\label{SubSec:SSS}

In this section, we take into account the physical scenario in which particles are accreted from a reservoir in the asymptotic region. We also recall that the gas molecules consist of identical massive and spinless particles. We further suppose that the gas is steady-state, spherically symmetric, and collisionless as described in Section~\ref{SubSec:ExSSSI}. Besides, we consider
an isotropic reservoir and assume that it is in thermodynamic equilibrium
after undergoing some physical process. Thus, the state of the gas can be
described by an equilibrium distribution function with a given temperature $%
T>0$ in the asymptotic region. It is worth mentioning that since we neglected collisions between the gas particles, the gas is not in local thermodynamic equilibrium at a finite radius.\footnote{See Ref.~\cite{pRoS17} for a discussion regarding this point.}  Therefore, $T$ should not be interpreted as a local temperature and should be only considered as an asymptotic parameter.

By considering the assumptions mentioned above, the function $f_\infty$ in Eq.~(\ref{Eq:SSSSIG}) takes the Maxwell-J\"uttner form~\cite{fJ11a,fJ11b} 
\begin{equation}
f_\infty(\varepsilon) = \alpha e^{-z\varepsilon},\qquad
z := \frac{mc^2}{k_B T},  \label{Eq:fEquilibrium}
\end{equation}
where $\alpha > 0$ is a normalization constant and $z$ is proportional to the inverse temperature. From this point on, we reintroduce the speed of light $c$ for convenience. It is notable to mention that for most astrophysical systems, the rest energy of the particles is much larger than the thermal energy, such that $z\gg 1$. For instance, this parameter$\ $is
of the order of $10^{9}$ for gas accreted from the interstellar medium~\cite{Shapiro-Book}. However, under extreme conditions (early universe, inner regions of neutron stars, photon gas limit), the ultrarelativistic regime $z\to 0$ is also important.

In the following, we substitute Eq.~(\ref{Eq:fEquilibrium}) into equations~(\ref{Eq:nepsp}) and (\ref{Defsab}--\ref{Dexp}), and we discuss the physical content of the resulting particle current
density and energy-momentum-stress tensor. In the asymptotic limit, the gas behaves like an isotropic ideal fluid, see Eq.~(\ref{Eq:JTInfty}). The choice~(\ref{Eq:fEquilibrium}) implies $f_\infty'(\varepsilon) = -z f_\infty(\varepsilon)$ and hence Eq.~(\ref{Eq:povern}) (with $m$ replaced with $mc^2$) leads to
\begin{equation}
p_\infty = n_\infty k_B T,
\end{equation}
which is the ideal gas equation of state. Furthermore, one has the identities
\begin{equation}
\mathcal{W}_{0,l}[f_\infty] = 4\pi m^2\alpha\frac{1\cdot 3\cdots (2l+1)}{z^{l+1}} K_{l+1}(z),\qquad l=0,1,2,\ldots
\end{equation}
with $K_l(z)$ denoting the modified Bessel function of the second kind (see for instance chapter 10 in Ref.~\cite{DLMF}), such that Eq.~(\ref{Eq:nepsp}) leads to
\begin{equation}
n_{\infty }(z) = 4\pi \alpha m^{3}\frac{K_{2}(z)}{z},
\qquad
\rho_\infty(z) = 4\pi \alpha m^{4} c^2\left[ \frac{K_{1}(z)}{z}+3\frac{K_{2}(z)}{z^{2}}\right] ,\qquad 
p_{\infty }(z) = 4\pi \alpha m^{4} c^2\frac{K_{2}(z)}{z^{2}}.
\label{n_infty}
\end{equation}

At the horizon, the state of the gas is described by the expressions in Eq.~(\ref{Eq:ObsHorizon}) where one replaces $f_\infty(\varepsilon)$ with $\alpha e^{-z\varepsilon}$, and one obtains the particle density $n_{h}$, energy density $\rho _{h}$, radial pressure $p_{rad,h}$, and tangential pressure $p_{tan,h}$, as described at the end of Section~\ref{SubSec:Horizon}.

To get further insight into the accretion process, in the following, we evaluate the observables in both the low temperature ($z\to\infty$) and the high temperature ($z\to 0$) limits.

\subsubsection{Low temperature limit}

We recall that $z$ is very large
in most ordinary astrophysical systems, hence it is worthwhile to take the limit $z\rightarrow
\infty$ in the expressions we have found so far. The details of this calculation are given in App.~\ref{App:Limits}, and here we directly provide the result:
\begin{eqnarray}
\lim\limits_{z\rightarrow \infty }\frac{\rho _{h}}{n_{h}} &=&\frac{mc^{2}}{2\sqrt{1+\frac{\Lambda ^{2}}{%
2}}}\left( 1+\frac{\Lambda ^{2}}{2}+\sqrt{1+\Lambda^2+\frac{\Lambda^4}{3}} \right) ,
\label{Eq:nH}\\
\lim\limits_{z\rightarrow \infty }\frac{p_{rad,h}}{n_{h}} &=&\frac{mc^{2}}{2\sqrt{1+\frac{\Lambda ^{2}}{2
}}}\left( -1-\frac{\Lambda ^{2}}{2}+\sqrt{1+\Lambda^2+\frac{\Lambda^4}{3}} \right) ,  
\label{Eq:pradH}\\
\lim\limits_{z\rightarrow \infty }\frac{p_{tan,h}}{n_{h}} &=&\frac{mc^{2}}{\sqrt{1+\frac{\Lambda ^{2}}{2
}}}\left( \frac{\Lambda }{2}\right) ^{2},
\label{Eq:ptanH}
\end{eqnarray}
where $\Lambda := \lambda _{c}\left( 1\right)/r_h$ and we recall that $L_c(E=m) = m\lambda _{c}\left( 1\right)$ is the critical total angular momentum for the energy corresponding to marginally bound orbits, that is, $r_{\max }(m\lambda _{c}\left( 1\right) )=r_{mb}$ (see the proof of Lemma \ref{Lem:RangeEmin}). Besides, the relations~(\ref{Eq:nH}--\ref{Eq:ptanH}) reduce to the corresponding ones for the standard Schwarzschild black hole in the limit $\Lambda=2$, as it should be (see Eqs. (83-85) in \cite{pRoS16}).

Moreover, using the results from App.~\ref{App:Limits}, one finds the following expressions for the four-velocity and the time- and spacelike eigenvectors $e_{0}$ and $e_1$ of $(T^{\mu
}{}_{\nu })$:
\begin{eqnarray}
u_{h} &=&\frac{1}{\sigma_h \sqrt{1+\frac{\Lambda ^{2}}{2}}}%
\left( \left[ 1+\frac{\sigma_h }{2}\left( 1+\frac{\Lambda ^{2}}{%
2}\right) \right] \frac{\partial }{\partial t}-\frac{\partial }{%
\partial r}\right) ,
\\
e_{0} &=&\frac{1}{\sigma_h}\left( 1+\Lambda^2+\frac{\Lambda^4}{3}\right)^{-1/4} \left( \left[ 1+\frac{\sigma_h}{2}\sqrt{1+\Lambda^2+\frac{\Lambda^4}{3}}\right] \frac{\partial }{\partial t} -
\frac{\partial }{\partial r}\right),
\\
e_{1} &=&\frac{1}{\sigma_h}\left( 1+\Lambda^2+\frac{\Lambda^4}{3}\right)^{-1/4} \left( \left[ 1-\frac{\sigma_h}{2}\sqrt{1+\Lambda^2+\frac{\Lambda^4}{3}}\right] \frac{\partial }{\partial t} -
\frac{\partial }{\partial r}\right),
\end{eqnarray}
which shows that $u_{h}$ and $e_{0}$ are not parallel. The fact that $p_{rad,h} < p_{tan,h}$ and that $u_h$ and $e_0$ do not point in the same direction for any value of $\Lambda > 0$ and $\sigma_h > 0$ imply that the gas does not behave as an isotropic perfect fluid at the horizon, as announced at the end of Section~\ref{Sec:Observables}.

Finally, the mass accretion rate $\dot{M}$, energy accretion rate $\dot{\mathcal{E}}$, and compression ratio $n_{h}/n_{\infty }$ of the gas particles in the low temperature limit satisfy (see App.~\ref{App:Limits})
\begin{equation}
\lim\limits_{z\rightarrow \infty }\frac{1}{\sqrt{2\pi z}}\frac{\dot{M}}{%
n_{\infty }}=\Lambda^{2}r_h^2{}mc,\qquad
\lim\limits_{z\rightarrow \infty }\frac{1}{\sqrt{2\pi z}}\frac{\dot{\mathcal{E}}}{%
n_{\infty }}=\Lambda^{2}r_h^2{}mc^3,\qquad
\lim\limits_{z\rightarrow \infty }\frac{1}{\sqrt{2\pi z}}\frac{n_{h}}{%
n_{\infty }}=\frac{\Lambda ^{2}}{4\pi}\sqrt{1+\frac{\Lambda ^{2}}{%
2}},
\label{MdotLTL}
\end{equation}
which shows that for large values of $z$ they scale as $\sqrt{2\pi z} = \sqrt{2\pi m c^2/(k_B T)}$ (This factor is equal to the thermal wavelength divided by the reduced Compton wavelength of the gas particles). It is worthwhile to note that the corresponding quantities in the Michel model~\cite{fM72,ZelNovik-Book,Shapiro-Book,eCoS15a}, that describes the spherical steady-state
accretion of a polytropic perfect fluid, are larger than $\dot{M}$ and $n_{h}/n_{\infty}$ by a factor of $z$. Moreover, $\dot{M}/n_\infty$ and $n_h/n_\infty$ in Eq.~(\ref{MdotLTL}) reduce to the corresponding ones for the standard Schwarzschild black hole in the limit $\Lambda=2$, as we expected (see Eq. (87) in \cite{pRoS16}).

\subsubsection{High temperature limit}

Next, we analyze the ultra-relativistic limit $z\to 0$. Taking into account the arguments in App.~\ref{App:UltraRelativistic}, the energy density and the principal pressures on the horizon behave as
\begin{eqnarray}
\lim\limits_{z\rightarrow 0}z\frac{\rho _{h}}{%
n_{h}}&=&\frac{3mc^2 \zeta}{2\sqrt{2}}\left(1+\frac{2}{\sqrt{3}} \right), \label{Eq:nHz0}\\
\lim\limits_{z\rightarrow 0}z\frac{p_{rad,h}}{n_{h}}&=&\frac{3mc^2 \zeta}{2\sqrt{2}
}\left(-1+\frac{2}{\sqrt{3}} \right),
\label{Eq:pradHz0}\\
\lim\limits_{z\rightarrow 0}z\frac{p_{tan,h}}{n_{h}} &=&\frac{3mc^2 \zeta}{2\sqrt{2}},
\label{Eq:ptanHz0}
\end{eqnarray}
and thus they diverge as $z^{-1} = k_B T/(m c^2)$ for $T\to \infty$. In contrast, recall that in the low-temperature limit these quantities are constant. Here, $\zeta:=\tilde{\zeta}/r_h$ where we recall that $\tilde{\zeta}$ was defined in Eq.~(\ref{Eq:zetatilde}) and depends on the explicit form of the metric functions and properties of the underlying spacetime. Furthermore, in this case, the four-velocity and the time- and spacelike eigenvectors $e_{0}$ and $e_{1}$ of $(T^{\mu}{}_{\nu })$ are given by 
\begin{eqnarray}
u_{h} &=&\frac{\sqrt{2}}{\sigma_h \zeta}%
\left( \left[ 1+\frac{\sigma_h \zeta^2}{4} \right] \frac{\partial }{\partial t}-\frac{\partial }{%
\partial r}\right) ,\qquad  \\
e_{0} &=&\frac{3^{\frac{1}{4}}}{\sigma_h\zeta} \left( \left[1+\frac{\sigma_h \zeta^2}{2\sqrt{3}} \right] \frac{\partial }{\partial t}-\frac{\partial }{\partial r}\right) ,\\
e_{1} &=&\frac{3^{\frac{1}{4}}}{\sigma_h\zeta} \left( \left[1-\frac{\sigma_h \zeta^2}{2\sqrt{3}} \right] \frac{\partial }{\partial t}-\frac{\partial }{\partial r}\right) ,
\end{eqnarray}%
which shows that $u_{h}$ and $e_{0}$ are not parallel again. Finally, the mass and energy accretion rates and
compression ratio of the ultra-relativistic gas read (see App.~\ref{App:UltraRelativistic})
\begin{equation}
\lim\limits_{z\rightarrow 0}\frac{\dot{M}}{n_{\infty
}}=\pi \zeta^2r_h^2 m c ,\qquad
\lim\limits_{z\rightarrow 0}z\frac{\dot{\mathcal{E}}}{n_{\infty
}}=3\pi \zeta^2r_h^2 m c^3 ,\qquad
\lim\limits_{z\rightarrow 0}\frac{n_{h}}{n_{\infty }}%
=\sqrt{2}\left(\frac{\zeta}{2}\right)^3,
\label{MdotHTL}
\end{equation}
which shows that $\dot{M}/n_\infty$ and $n_h/n_\infty$ are constant, and unlike the low-temperature limit, they do not scale with the $z$ parameter, whereas $\dot{\mathcal{E}}/n_\infty$ diverges as $z^{-1} = k_B T/(mc^2)$. In addition, one can verify that $\dot{M}/n_\infty$ and $\dot{\mathcal{E}}/n_\infty$ given in Eq.~(\ref{MdotHTL}) reduce to the corresponding relations for the Schwarzschild solutions for $\zeta=3\sqrt{3}/2$ (see Eqs. (14-15) in \cite{pRoS17}). As for the Schwarzschild case, the different behavior between $\dot{M}/n_\infty$ and $\dot{\mathcal{E}}/n_\infty$ can be explained by the fact that as the temperature increases, a lower fraction of the particles is accreted; however $\dot{\mathcal{E}}/n_\infty$ keeps growing with $T$ since these particles have higher energy.

It is worthwhile to mention that for arbitrary values of $z$, one can perform the integrations presented in Eqs.~(\ref{Eq:JainSS}--\ref{Eq:TthetathetascatSS}) numerically to obtain the energy density, principal pressures, accretion rate, and compression ratio at an arbitrary point $r\in[r_h,\infty)$ in a similar manner as performed in~\cite{pRoS17} for the Schwarzschild black hole and in~\cite{aCpM20} for the Reissner-Nordstr\"om black hole. Here, we have only considered the cases of extreme temperature at the horizon and at infinity in order to find analytic formulae for these quantities.

To reach further physical conclusions from our obtained results, we need to know the explicit form of the function $\sigma(r)$ as well as the values of $\Lambda$, $\zeta$, and $r_{h}$. This matter will be investigated
in the upcoming subsection for a couple of black hole spacetimes in order to show the applications of our generic formulae.

\subsection{Accretion into black hole spacetimes in general relativity and beyond}
\label{SubSec:Examples}

\begin{figure*}[tbh]
\centering
\includegraphics[width=0.49\textwidth]{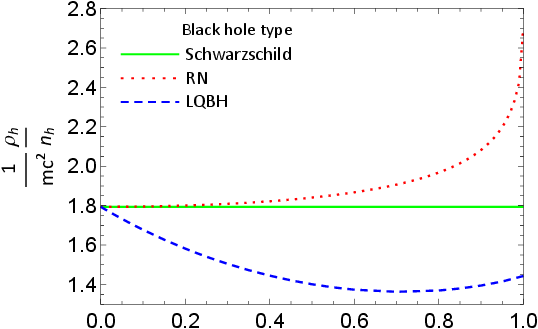} \includegraphics[width=0.49%
\textwidth]{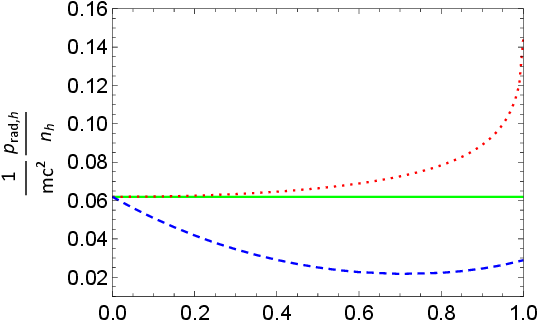}\newline
\includegraphics[width=0.49\textwidth]{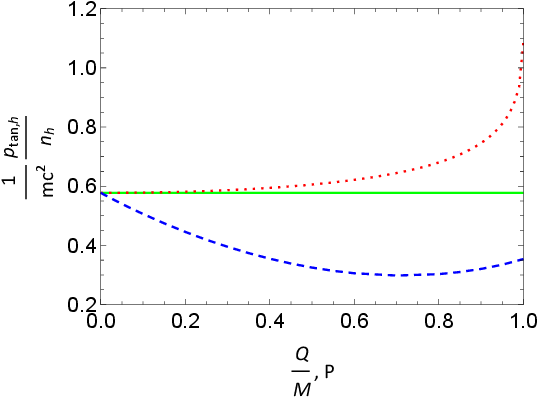} \includegraphics[width=0.49%
\textwidth]{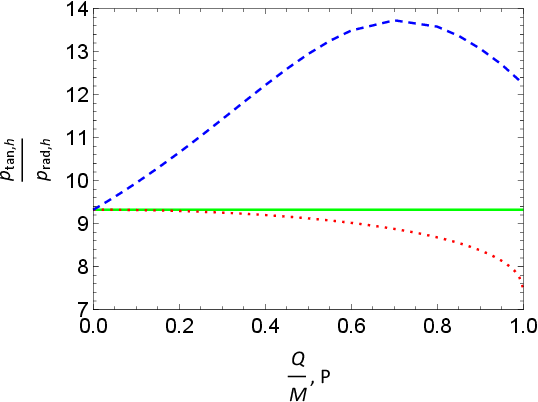}
\caption{The behavior of the energy density and principal pressures at the event horizon as functions of the deformation parameters in the low temperature limit. The values corresponding to the Schwarzschild black hole are included for reference.}
\label{rhoLowT}
\end{figure*}
\begin{figure*}[tbh]
\centering
\includegraphics[width=0.49%
\textwidth]{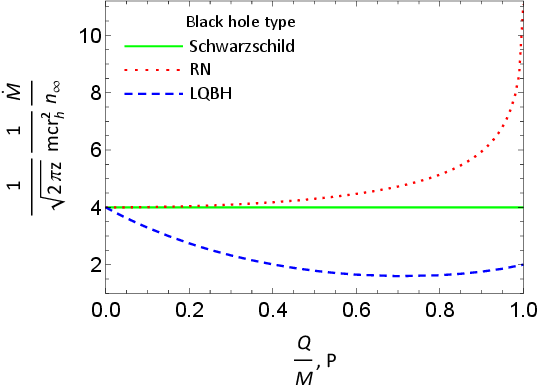}
\includegraphics[width=0.49%
\textwidth]{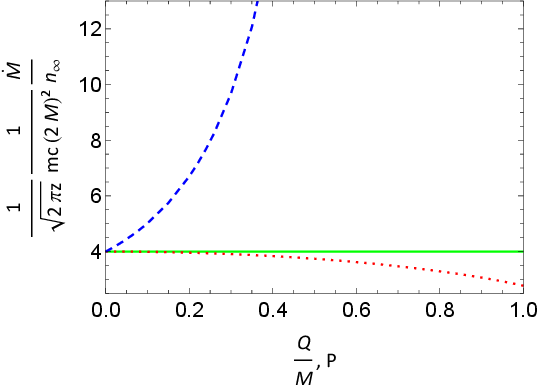}
\caption{The behavior of the mass accretion rate as a function of the deformation parameters in the low temperature limit. $\dot{M}$ is normalized by the square of the event horizon radius (left panel) and by the total black hole mass (right panel). The values corresponding to the Schwarzschild black hole are included for reference.}
\label{accLowT}
\end{figure*}
\begin{figure*}[tbh]
\centering
\includegraphics[width=0.49\textwidth]{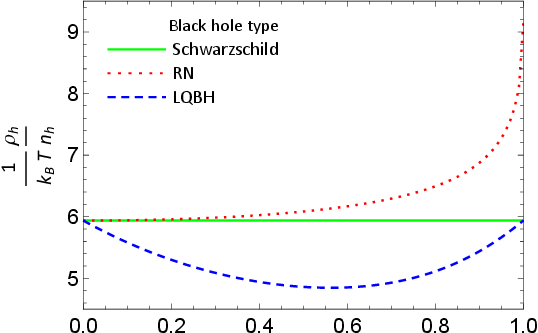} \includegraphics[width=0.49
\textwidth]{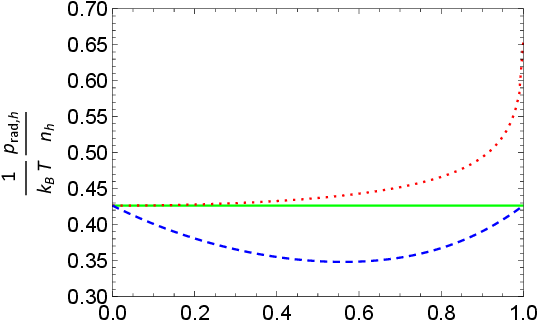}\newline
\includegraphics[width=0.49\textwidth]{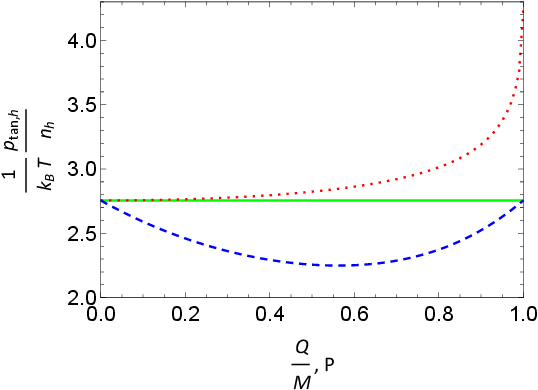}
\includegraphics[width=0.49\textwidth]{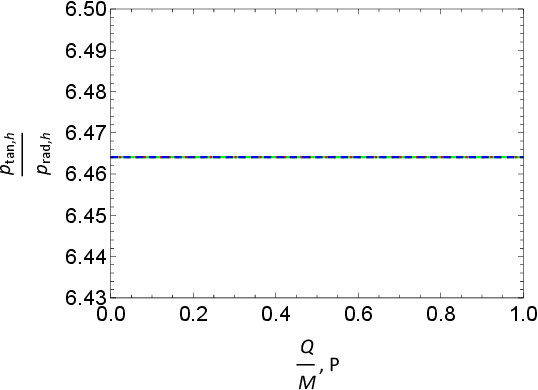}
\caption{The behavior of the energy density and principal pressures at the event horizon as functions of the deformation parameters in the high temperature limit. Note that these quantities scale like $k_B T$ for high temperatures. Also observe that the ratio $p_{tan,h}/p_{rad,h}$ is constant in the high temperature limit, as predicted by Eqs.~(\ref{Eq:pradHz0},\ref{Eq:ptanHz0}).}
\label{rhoHiT}
\end{figure*}
\begin{figure*}[tbh]
\centering
\includegraphics[width=0.49%
\textwidth]{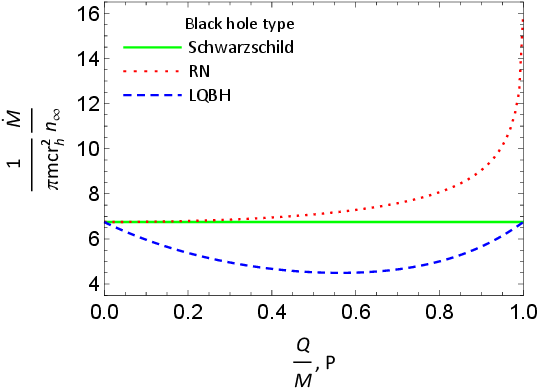}
\includegraphics[width=0.49%
\textwidth]{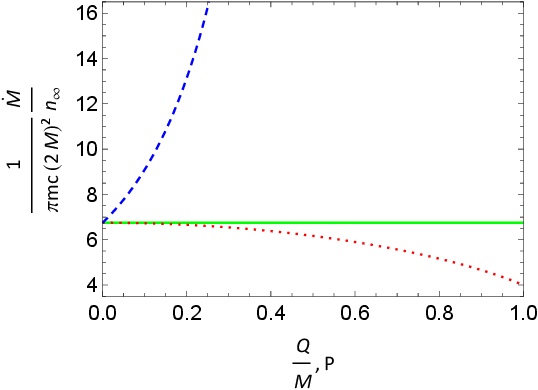}
\caption{The behavior of the mass accretion rate as a function of the deformation parameters in the high temperature limit. $\dot{M}$ is normalized by the square of the event horizon radius (left panel) and by the total black hole mass (right panel).}
\label{accHiT}
\end{figure*}
\begin{figure*}[tbh]
\centering
\includegraphics[width=0.49%
\textwidth]{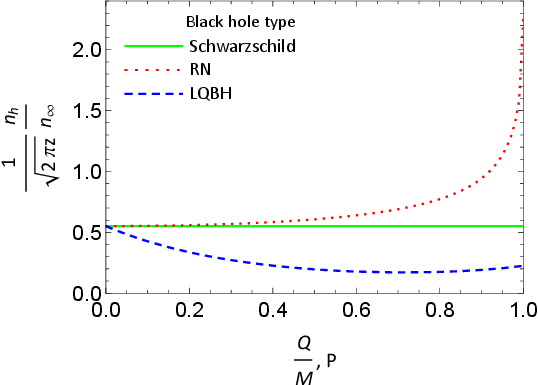}
\includegraphics[width=0.49%
\textwidth]{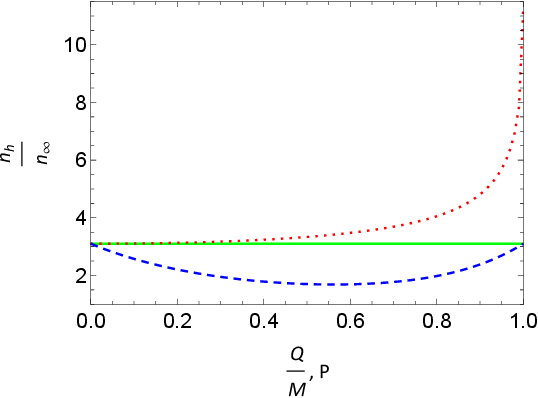}
\caption{The compression ratio as a function of the deformation parameters $Q/M$ and $P$ in the low (left panel) and high temperature (right panel) limits.}
\label{accLandH}
\end{figure*}

In this last subsection, in order to show the application of our generic formulae, we explore two special black hole spacetimes as concrete examples. The first example considers the Reissner–Nordstr\"om (RN) black holes which constitute the most general spherically symmetric black hole solution of the Einstein-Maxwell system. As the second example, we take into account a regular loop quantum black hole (LQBH) model that was proposed in Ref.~\cite{lM2010}. In both cases, we investigate the effects of the free parameters characterizing the black hole spacetimes on the observable quantities and explore deviations from the Schwarzschild black hole case.  

The RN black hole spacetime possesses an electric charge parameter $Q$ in addition to the black hole mass $M$, where $Q^2\leq M^2$. The corresponding metric functions are given by
\begin{equation}
\sigma_{RN}(r)=1, \quad \psi_{RN}(r)=1-\frac{2M}{r}+\frac{Q^2}{r^2}=\frac{1}{r^2}(r-r_{h,RN})(r-r_{c,RN}),
\label{RNmetric}
\end{equation}
and the event and Cauchy horizons are located at $r_{h,RN}=M+\sqrt{M^2-Q^2}$ and $r_{c,RN}=M-\sqrt{M^2-Q^2}$, respectively. For simplicity, in what follows, we restrict ourselves to the subextremal case $Q^2 < M^2$.

The loop quantum gravity (LQG) corrected metric functions are given by
\cite{lM2010}
\begin{equation}
\sigma_{LQ}(\bar{r})=\left(1+\frac{r_0}{\bar{r}}\right)^2, \qquad
\psi_{LQ}(\bar{r})=\frac{\bar{r}^4(\bar{r}-r_{h,LQ})(\bar{r}-r_{c,LQ})}{(\bar{r}^4+a_0^2)(\bar{r}+r_0)^2}, \label{LQBHmetric}
\end{equation}
where the relation between the areal radius $r$ and the radial coordinate $\bar{r}$ is given by
\begin{equation}
r^2 = \bar{r}^2+\frac{a_0^2}{\bar{r}^2},\qquad \bar{r} > 0.
\end{equation}
For this black hole, the event horizon is located at $\bar{r} = r_{h,LQ}=2M_0/(1+P)^{2}$ and the Cauchy horizon at $\bar{r} = r_{c,LQ}=2M_0 P^{2}/(1+P)^{2}$ where the polymeric function $P=\left(
\sqrt{1+\epsilon ^{2}}-1\right) /\left( \sqrt{1+\epsilon ^{2}}+1\right)$ stems from the geometric quantum effects of LQG. Besides, $r_{0}=\sqrt{r_{h,LQ}\times r_{c,LQ}}=2M_0 P/(1+P)^{2}$, $a_0$ is related to the minimum area gap of LQG by $a_0=a_{\min}/(8 \pi)$, $M_0$ is a mass parameter related to the total mass through $M = M_0/(1+\epsilon^2)$, and $\epsilon$ denotes the product of the
Immirzi parameter $\gamma$ and the polymeric parameter $\delta$ satisfying $\epsilon =\gamma \delta\ll 1$. Therefore, the LQG corrected metric has three parameters: $M_0$, $\epsilon$, and $a_0$. Note that $r_{h,LQ} > r_{c,LQ} > 0$ because $0 < P < 1$.

Since $a_0$ is microscopic and in this work we are only interested in the region outside the event horizon (such that $\bar{r} \geq r_{h,LQ}\gg \sqrt{a_0}$), we set $a_0 = 0$ in this study. This simplification implies that $\bar{r} = r$ and $\psi_{LQ}(r) = (r -r_{h,LQ})(r - r_{c,LQ})/(r+r_0)^2$ in Eq.~(\ref{LQBHmetric}), and we take the remaining free parameters to be $M_0$ and $P$ (instead of $M_0$ and $\epsilon$). Note that $P = \mathcal{O}(\epsilon^2)$ is small in LQG, and a recent work based on shadow observations obtains the restriction $P < 0. 082$~\cite{Pbound}. Nevertheless, we shall explore the whole range $0 < P < 1$ for completeness.

In the metric functions~(\ref{RNmetric}-\ref{LQBHmetric}), 
the free parameters $Q$ and $P$ indicate deviations from
the standard Schwarzschild black holes and they reduce to the
Schwarzschild solutions by taking the limits $Q\to0 $ and $ P\to0$. Moreover, as verified in App.~\ref{App:Validity}, these metrics satisfy the assumptions (i)-(iv) listed in Section~\ref{SubSec:GenericBH}, which underlie our analysis.

Figures~\ref{rhoLowT},\ref{accLowT},\ref{rhoHiT},\ref{accHiT}, and \ref{accLandH} show the behavior of the relevant quantities in Eqs.~(\ref{Eq:nH},\ref{Eq:pradH},\ref{Eq:ptanH},\ref{MdotLTL},\ref{Eq:nHz0},\ref{Eq:pradHz0},\ref{Eq:ptanHz0},\ref{MdotHTL}) as a function of the dimensionless deformation parameters $Q/M$ and $P$. To do these plots, we have calculated $r_h$, $\Lambda$, and $\zeta$ that appear in these expressions (see App.~\ref{App:Lc} for details). Interestingly, with respect to the standard Schwarzschild case, the energy density, principal pressures (when normalized by the rest energy density), as well as the compression ratio are enhanced for RN and suppressed for the LQBH. Note that these quantities depend on the parameters $\Lambda$ and $\zeta$ in the low- and high-temperature limits which in the RN (LQBH) case are larger (smaller) than the standard Schwarzschild case. When normalized by the square of the event horizon radius, the mass accretion rate shows a similar qualitative behavior. However, in view of astrophysical applications, it might be more useful to normalize the accretion rate with respect to the square of the total mass instead of $r_h^2$. The corresponding results are shown in the right panels of Figs.~\ref{accLowT} and \ref{accHiT}. As can be seen from these figures, with respect to this normalization the accretion rate is suppressed in the RN case, whereas it can be enhanced by a large factor in the LQBH case. This qualitative difference between the two normalizations can be partially understood by considering that
\begin{equation}
\frac{r_h}{2M} = \left\{ \begin{array}{cc} \frac{1}{2}\left[ 1 + \sqrt{1 - \left( \frac{Q}{M} \right)^2} \right], & \hbox{for RN},\\
 \frac{1}{(1-P)^2}, & \hbox{for LQBH,}
 \end{array} \right. 
\end{equation}
which is smaller (larger) than one in the RN (LQBH) case and diverges when $P\to 1$. Finally, we would like to mention that our results in the RN case are in qualitative agreement with the analysis performed in~\cite{aCpM20}, where the physical quantities were plotted for values of $z$ in the range $1\leq z\leq 30$.


\section{Conclusions}

\label{Sec:Conclusions} 

We have extended the spherical accretion problem of a collisionless kinetic gas into the standard Schwarzschild black hole to generic static and spherically symmetric black hole spacetimes. This extension allows us to explore the effects of free parameters characterizing black hole spacetimes beyond Einstein vacuum gravity on the accretion process. To do so, we identified a large class $\mathcal{C}$ of static and spherically symmetric black holes whose metric is asymptotically flat, has a positive total mass, a regular Killing horizon, and satisfies some monotonicity conditions. As we have shown, these conditions guarantee that the effective potential $V_{m,L}$ associated with the radial motion of the gas particles in these generic spacetimes has the same qualitative behavior as in the standard Schwarzschild case. As in previous work, we assumed that the gas is dilute enough in order to ignore its self-gravity and collisions between the gas particles, that the accretion rate is so small that the spacetime background remains static over large time scales, and that the gas does not interact directly with other matter fields. Under these assumptions, the problem reduces to solving the collisionless Boltzmann equation for the one-particle distribution function on a fixed background spacetime. 

To solve this problem, we followed the Hamiltonian formalism and performed the computations on the cotangent bundle $T^{\ast }\mathcal{M}$ associated with a fixed spacetime manifold $(\mathcal{M},g)$ within the class $\mathcal{C}$. We introduced horizon-penetrating coordinates in order to facilitate the interpretation of the spacetime observables at the horizon. The structure of the invariant subsets $\Gamma _{m,E,L_{z},L }\subset T^{\ast }\mathcal{M}$ defined by the conserved quantities associated with the geodesic motion was explored and the qualitative behavior of the effective potential $V_{m,L}$ was
analyzed. For the accretion problem under consideration in this article, we limited the calculations to the invariant submanifold $\Gamma_{accr}$ of the relativistic phase space in order to neglect contributions from particles on bound orbits or particles emanating from the white hole region. One of our main results (see Theorem~\ref{Thm:Main}) consists in presenting the most general collisionless distribution function on $\Gamma_{accr}$ in terms
of the symplectic coordinates $\left( Q^{\mu },P_{\mu }\right)$ for the case of generic black holes within the class $\mathcal{C}$. For the accretion problem considered in this article, one further assumes that asymptotically, the gas behaves as an isotropic perfect fluid which is at rest. This fully determines the distribution function on $\Gamma_{accr}$ up to the function $f_\infty$ which describes the energy distribution at infinity, see Eq.~(\ref{Eq:SSSSIG}).

A fundamental step in our analysis was to show that there is an energy-dependent critical angular
momentum $L _{c}(E)$ such that infalling particles from infinity are absorbed by the black hole when $L < L _{c}(E)$ and scattered off the centrifugal barrier when $L > L _{c}(E)$. When applied to the accretion problem, the former particles contribute to the accretion rate but not to the physical quantities in the asymptotic region due to the fact that the observables associated with these particles decay to zero as $r\to \infty$. Conversely, the scattered particles do yield nontrivial observables in the asymptotic region; however they contribute neither to the accretion rate nor to the physical quantities on the horizon since they carried enough angular momentum to be scattered off the effective potential. 

Furthermore, we have calculated the physical observables relevant to the accretion process, namely the particle current density and the energy-momentum-stress tensor, and we have compared their properties obtained from the simple distribution function $f_{\infty}$ depending only on the energy with those obtained from an isotropic perfect fluid. Whereas we have shown that the gas behaves as an ideal isotropic fluid at infinity, it may behave rather differently near the horizon.

Next, we considered the special case in which $f_\infty$ is given by a Maxwell-J\"uttner distribution function, corresponding to a gas in thermodynamic equilibrium with a given temperature $T$ at rest in the asymptotic region. For such configuration, we computed the particle flux $j_{n}$ and energy flux $j_{\varepsilon }$ passing through a sphere of constant radius outside the event horizon, from which the mass and energy accretion rates $\dot{M}$ and $\dot{\mathcal{E}}$ can be determined. Further, we calculated the particle density $n_{\infty }$, energy density $\rho_\infty$, and pressure $p_{\infty }$ at infinity and verified that they satisfy the ideal gas equation $p_\infty = n_\infty k_B T$. Besides, we computed the compression ratio $n_{h}/n_{\infty }$ as well as the particle density $n_{h}$, energy density $\rho _{h}$, radial pressure $p_{rad,h}$, and tangential pressure $p_{tan,h}$ of the gas on the horizon. In particular, we analyzed the behavior of these quantities in both the low and high temperature limits. In these limits, it was explicitly shown that the tangential pressure of the gas is always larger than the radial one on the horizon (see Eqs.~(\ref{Eq:pradH}--\ref{Eq:ptanH}) and Eqs.~(\ref{Eq:pradHz0}--\ref{Eq:ptanHz0})),
implying that, unlike at infinity, the gas does not behave like an isotropic perfect fluid at the horizon. Interestingly, the physical observables at the horizon only depend on the generic metric functions through the quantities $r_h$ (the event horizon radius), $\sigma_h$, and $L_c(E)$ (which only depends on $\chi(r)$). Remarkably, the mass and energy accretion rates  are independent of $r_h$ and $\sigma_h$, and hence they only depend on the curvature of the spacetime through the behavior of the critical angular momentum $L_c(E)$ (see Eqs.~(\ref{Eq:MassAccretionRate},\ref{Eq:EnergyAccretionRate})). Regarding $L_c(E)$, in the low temperature limit, only its value for the lowest energy $E=m$ is relevant, whereas in the high temperature limit, only its asymptotic slope $\tilde{\zeta}$ is required (see Eq.~(\ref{Eq:zetatilde})). Although we have only explored the limiting cases of high- and low-temperature limits at the horizon and at infinity, the general formulae we have provided serve as a starting point for a numerical study of the behavior of the physical observables for arbitrary temperatures and locations.

Finally, in order to show the application of our general formulae for generic static and spherically symmetric black hole spacetimes, we considered two concrete examples. The first one is the RN solution of the Einstein-Maxwell system and the second one is a LQBH model. We investigated the effects of the electric charge and the deformation parameter $P$ on the observables at the horizon and the accretion rate and compression ratio, and compared the results with those of the standard Schwarzschild black hole. Compared to the latter, we have found that the energy density, principal pressures (when normalized by the rest energy density), as well as the compression ratio are enhanced for RN and suppressed for the LQBH. When normalized by $r_h^2$, the mass accretion rate shows a similar qualitative behavior. In contrast, when normalized by the square of the total mass, the accretion rate is suppressed in the RN case and enhanced in the LQBH case. Although this enhancement can in principle be arbitrarily large, for the range of validity $0 < P < 0.082$ determined in Ref.~\cite{Pbound} of the deformation parameter, it stays below $30\%$.

We hope the framework developed in this paper as well as our general formulae will be useful to explore the accretion problem in various other black hole spacetimes.


\acknowledgments

It is our pleasure to thank Thomas Zannias for fruitful and stimulating discussions. This work was partially supported by CIC Grant No.~18315 to Universidad Michoacana, by CONAHCyT Network Project No.~376127 ``Sombras, lentes y ondas gravitatorias generadas por objetos compactos astrof\'isicos", and by CONAHCyT-SNII. MM was also supported by CONAHCyT through Estancias
Posdoctorales por Mexico Convocatoria 2023(1) under the postdoctoral Grant No. 1242413.

\appendix

\section{Integral identities}

\label{App:integrals} 

In this appendix, we summarize some of the integral identities used in the
derivation of equations~(\ref{Eq:JainSS}--\ref{Eq:TthetathetascatSS}). By
introducing the shorthand notation $s_{1,2}:=\sqrt{\varepsilon
^{2}-U_{\lambda _{1,2}}(r)}$ and for any $0\leq \lambda _{1}<\lambda
_{2}\leq \lambda _{max}$, we have 
\begin{eqnarray}
\int\limits_{\lambda _{1}}^{\lambda _{2}}\frac{\lambda d\lambda }{\sqrt{%
\varepsilon ^{2}-U_{\lambda }(r)}} &=&-\frac{r^{2}}{\chi }(s_{2}-s_{1})=%
\frac{\lambda _{2}^{2}-\lambda _{1}^{2}}{s_{1}+s_{2}}, \\
\int\limits_{\lambda _{1}}^{\lambda _{2}}\pi _{r\pm }(r)\frac{\lambda
d\lambda }{\sqrt{\varepsilon ^{2}-U_{\lambda }(r)}} &=&\frac{\lambda
_{2}^{2}-\lambda _{1}^{2}}{\sigma \psi }\left( \frac{\varepsilon (1-\sigma
\psi )}{s_{1}+s_{2}}\pm \frac{1}{2}\right) , \label{Eq:2ndIdentity} \\
\int\limits_{\lambda _{1}}^{\lambda _{2}}\pi _{r\pm }^{2}(r)\frac{\lambda
d\lambda }{\sqrt{\varepsilon ^{2}-U_{\lambda }(r)}} &=&\sigma ^{2}\frac{%
\lambda _{2}^{2}-\lambda _{1}^{2}}{s_{1}+s_{2}}\left[ \left( \frac{%
\varepsilon (1-\sigma \psi )\pm s_{1}}{\chi }\right) \left( \frac{%
\varepsilon (1-\sigma \psi )\pm s_{2}}{\chi }\right) +\frac{1}{3}\left( 
\frac{s_{2}-s_{1}}{\chi }\right) ^{2}\right] , \label{Eq:3ndIdentity} \\
\int\limits_{\lambda _{1}}^{\lambda _{2}}\lambda ^{2}\frac{\lambda d\lambda 
}{\sqrt{\varepsilon ^{2}-U_{\lambda }(r)}} &=&\frac{1}{3}\frac{\lambda
_{2}^{2}-\lambda _{1}^{2}}{(s_{1}+s_{2})^{2}}\left[ \lambda
_{1}^{2}(s_{1}+2s_{2})+\lambda _{2}^{2}(s_{2}+2s_{1})\right] .
\end{eqnarray}%
In particular, the identities~(\ref{Eq:2ndIdentity}--\ref{Eq:3ndIdentity}) for $\pi _{r}=\pi _{r-}(r)$ 
can be also written as
\begin{eqnarray}
\int\limits_{\lambda _{1}}^{\lambda _{2}}\pi _{r-}(r)\frac{\lambda d\lambda 
}{\sqrt{\varepsilon ^{2}-U_{\lambda }(r)}} &=&-\frac{\lambda
_{2}^{2}-\lambda _{1}^{2}}{s_{1}+s_{2}}\left( \varepsilon -\frac{\sigma }{2}%
\frac{1+\frac{\lambda _{1}^{2}}{r^{2}}}{\varepsilon +s_{1}}-\frac{\sigma }{2}%
\frac{1+\frac{\lambda _{2}^{2}}{r^{2}}}{\varepsilon +s_{2}}\right) , \\
\int\limits_{\lambda _{1}}^{\lambda _{2}}\pi _{r-}^{2}(r)\frac{\lambda
d\lambda }{\sqrt{\varepsilon ^{2}-U_{\lambda }(r)}} &=&\sigma ^{2}\frac{%
\lambda _{2}^{2}-\lambda _{1}^{2}}{s_{1}+s_{2}}\left[ \left( \frac{%
\varepsilon }{\sigma }-\frac{1+\frac{\lambda _{1}^{2}}{r^{2}}}{\varepsilon
+s_{1}}\right) \left( \frac{\varepsilon }{\sigma }-\frac{1+\frac{\lambda
_{2}^{2}}{r^{2}}}{\varepsilon +s_{2}}\right) +\frac{1}{3r^{4}}\left( \frac{%
\lambda _{2}^{2}-\lambda _{1}^{2}}{s_{1}+s_{2}}\right) ^{2}\right] ,
\end{eqnarray}%
which shows that the corresponding expressions are regular on the event
horizon $r=r_{h}$.


\section{Low-temperature limit}

\label{App:Limits} 

In this appendix, we give some details regarding the necessary calculations
for analyzing the low-temperature limit $z=m c^2/(k_B T) \rightarrow \infty $ in
section~\ref{SubSec:SSS}. We first calculate the ratio between the mass
accretion rate and the particle density at infinity as follows 
\begin{equation}
\frac{\dot{M}}{n_{\infty }}=\frac{m|j_{n}|}{n_{\infty }}.
\end{equation}
The particle density at infinity~(\ref{n_infty}) can be written as 
\begin{equation}
n_{\infty }=4\pi \alpha m^{3}\int\limits_{1}^{\infty }\varepsilon \sqrt{%
\varepsilon ^{2}-1}e^{-z\varepsilon }d\varepsilon ,
\label{nInftyB2}
\end{equation}%
and by using the mass accretion rate~(\ref{Eq:MassAccretionRate}), we obtain
\begin{equation}
\frac{\dot{M}}{n_\infty} =
 \pi m\frac{\int\limits_{1}^{\infty }\lambda _{c}^{2}(\varepsilon )e^{-z(\varepsilon
-1)}d\varepsilon }{\int\limits_{1}^{\infty }\varepsilon \sqrt{\varepsilon
^{2}-1}e^{-z(\varepsilon -1)}d\varepsilon } = \pi m\sqrt{z}\frac{\int\limits_{0}^{\infty }\lambda _{c}^{2}(1+\frac{x}{z}%
)e^{-x}dx}{\int\limits_{0}^{\infty }\left( 1+\frac{x}{z}\right) \sqrt{2x+%
\frac{x^{2}}{z}}e^{-x}dx},
\label{MdotOvernInf}
\end{equation}
where we have applied the variable substitution $\varepsilon \mapsto
x=z(\varepsilon -1)$ in the last step. Now, we divide both sides of the
relation by $\sqrt{z}$ and take the limit $z\rightarrow \infty$, which yields
\begin{equation}
\lim\limits_{z\rightarrow \infty }\frac{1}{\sqrt{z}}\frac{\dot{M}}{n_{\infty
}} 
 = \pi m\lambda_c^2(1)\frac{\int\limits_{0}^{\infty
}e^{-x}dx}{\int\limits_{0}^{\infty }\sqrt{2x}e^{-x}dx}
 = \sqrt{2\pi} m\lambda_c^2(1).
\end{equation}
Similarly, one can compute the energy accretion rate by employing Eqs.~(\ref{Eq:EnergyAccretionRate}) and (\ref{nInftyB2}) as
\begin{equation}
\lim\limits_{z\rightarrow \infty }\frac{1}{\sqrt{z}}\frac{\dot{\mathcal{E}}}{n_{\infty
}} = \sqrt{2\pi} m\lambda_c^2(1).
\end{equation}

Now, in order to obtain the particle density at the horizon, one can first
follow a similar computation to find the ratios $a/n_{\infty }$ and $b/n_{\infty }$ from
equations~(\ref{Defsab}) and (\ref{nInftyB2}) 
\begin{equation}
\lim\limits_{z\rightarrow \infty }\frac{1}{\sqrt{z}}\frac{a}{n_{\infty }}=\frac{\Lambda ^{2}}{4\sigma_h}\sqrt{\frac{2}{%
\pi }},\qquad
\lim\limits_{z\rightarrow \infty }\frac{1}{\sqrt{z}}\frac{b}{n_{\infty }}%
=\frac{\Lambda ^{2}}{8}\left( 1+\frac{\Lambda ^{2}}{2}\right) \sqrt{\frac{2}{\pi 
}}, 
\end{equation}%
where $\Lambda = \lambda_c(1)/r_h$. Then we use Eq.~(\ref{Eq:ObsHorizon1}) to find the compression ratio as
\begin{equation}
\lim\limits_{z\rightarrow \infty }\frac{1}{\sqrt{z}}\frac{n_{h}}{n_{\infty }}%
=\frac{\Lambda ^{2}}{2\sqrt{2\pi }}\sqrt{1+\frac{\Lambda
^{2}}{2}}.  \label{Eq:nHninfty}
\end{equation}
Next, we divide the expressions for $A$, $B$, $H$, $D$ in Eqs.~(\ref{Aexp})-(\ref{Dexp}) by $\sqrt{z}n_{\infty
}$ and take the limit $z\rightarrow \infty $ to obtain 
\begin{eqnarray}
\lim\limits_{z\rightarrow \infty }\frac{1}{\sqrt{z}}\frac{A}{n_{\infty }} &=&%
\frac{\Lambda ^{2}m}{4\sigma_h}\sqrt{\frac{2}{\pi }},\qquad
\lim\limits_{z\rightarrow \infty }\frac{1}{\sqrt{z}}\frac{B}{n_{\infty }}=%
\frac{\Lambda ^{2}m}{8}\left( 1+\frac{\Lambda ^{2}}{2}%
\right) \sqrt{\frac{2}{\pi }},
\\
\lim\limits_{z\rightarrow \infty }\frac{1}{\sqrt{z}}\frac{H}{n_{\infty }} &=&
\frac{\sigma_h \Lambda ^{2}m}{16}\left( 1+\Lambda^2+\frac{\Lambda^4}{3}\right) \sqrt{\frac{2}{\pi }},\qquad
\lim\limits_{z\rightarrow \infty }\frac{1}{\sqrt{z}}\frac{D}{%
n_{\infty }}=\frac{\Lambda ^{4}m}{16}\sqrt{\frac{2}{\pi }},
\end{eqnarray}
from which 
\begin{eqnarray}
\lim\limits_{z\rightarrow \infty }\frac{1}{\sqrt{z}}\frac{\rho_{h}}{n_{\infty }} &=& \frac{\Lambda ^{2}m}{8}\left[ 1+\frac{\Lambda ^{2}}{2}+\sqrt{1+\Lambda^2+\frac{\Lambda^4}{3}}\right] \sqrt{\frac{2}{\pi }},
\\
\lim\limits_{z\rightarrow \infty }\frac{1}{\sqrt{z}}\frac{p_{rad,h}}{n_{\infty
}} &=& \frac{\Lambda ^{2}m}{8}\left[ -1-\frac{\Lambda ^{2}}{2}+\sqrt{1+\Lambda^2+\frac{\Lambda^4}{3}}\right] \sqrt{\frac{2}{\pi }},
\\
\lim\limits_{z\rightarrow \infty }\frac{1}{\sqrt{z}}\frac{p_{tan,h}}{n_{\infty
}} &=& \left( \frac{\Lambda }{2}\right) ^{4}m\sqrt{\frac{2}{\pi }},
\end{eqnarray}
using Eq.~(\ref{Eq:ObsHorizon2}). With the help of these relations and equation~(\ref{Eq:nHninfty}), we can
find the energy density $\rho _{h}$, radial pressure $p_{rad,h}$, and
tangential pressure $p_{tan,h}$ at the horizon presented in equations~(\ref%
{Eq:nH},\ref{Eq:pradH},\ref{Eq:ptanH}).


\section{Ultra-relativistic limit}

\label{App:UltraRelativistic} 

In this appendix, we give some details regarding the ultra-relativistic limit $z=m c^2/(k_B T) \rightarrow 0 $. The ratio between the mass
accretion rate and the particle density at infinity is given by (\ref{MdotOvernInf})
\begin{equation}
\frac{\dot{M}}{n_{\infty }} = \pi m\frac{
\int\limits_{1}^{\infty }\lambda _{c}^{2}(\varepsilon)e^{-z\varepsilon}d\varepsilon }{\int\limits_{1}^{\infty }\varepsilon \sqrt{\varepsilon
^{2}-1}e^{-z\varepsilon}d\varepsilon }
= \pi m\frac{
\int\limits_{z}^{\infty } \frac{z^2}{x^2}\lambda _{c}^{2}(\frac{x}{z} ) x^2 e^{-x}dx}{\int\limits_{z}^{\infty } x\sqrt{ x^2 - z^2} e^{-x}dx},
\end{equation}
where we have applied the variable substitution $\varepsilon \mapsto
x=z\varepsilon$ in the last equality. Now, we take the limit $z\rightarrow 0$ to obtain the mass accretion rate in the ultra-relativistic limit as
follows
\begin{equation}
\lim\limits_{z\rightarrow 0}\frac{\dot{M}}{n_{\infty
}}= \pi m\tilde{\zeta} ^{2}\frac{\int\limits_{0}^{\infty
}x^2e^{-x}dx}{\int\limits_{0}^{\infty }x^2e^{-x}dx}= \pi m\zeta^2 r_h^2,
\end{equation}
where we have used the limit~(\ref{Eq:zetatilde}) and recall that $\zeta:=\tilde{\zeta}/r_h$. Besides, the energy accretion rate reads
\begin{equation}
\lim\limits_{z\rightarrow 0}z\frac{\dot{\mathcal{E}}}{n_{\infty
}}= 3 \pi m\zeta^2 r_h^2.
\end{equation}

Now, by following a similar computation, one can calculate the particle density at the horizon in the ultra-relativistic limit. We first obtain the ratios $a/n_{\infty }$ and $b/n_{\infty }$ using  Eqs.~(\ref{Defsab}) and (\ref{nInftyB2}) as 
\begin{equation}
\lim\limits_{z\rightarrow 0}\frac{a}{%
n_{\infty }}=\frac{\zeta ^{2}}{4\sigma_h}, \qquad
\lim\limits_{z\rightarrow 0}\frac{b}{n_{\infty }}%
=\frac{\zeta ^4}{16}.
\end{equation}
Then, we employ Eq.~(\ref{Eq:ObsHorizon1}) to get
\begin{equation}
\lim\limits_{z\rightarrow 0}\frac{n_{h}}{n_{\infty }}%
=\sqrt{2}\left(\frac{\zeta}{2}\right)^3.  \label{Eq:nHninftyz0limit}
\end{equation}

Next, we compute the energy density and principal pressures on the
horizon for the ultra-relativistic limit. To do so, we multiply the expressions for $A$, $B$, $H$, and $D$ given in Eqs. (\ref{Aexp})-(\ref{Dexp}) by $z/n_{\infty}$ and take the limit $z\rightarrow 0$ to obtain
\begin{eqnarray}
\lim\limits_{z\rightarrow 0}z\frac{A}{n_{\infty }} &=&\frac{3m \zeta^2}{4\sigma_h},\qquad
\lim\limits_{z\rightarrow 0}z\frac{B}{n_{\infty }}=\frac{3m \zeta^4}{16}, \qquad
\lim\limits_{z\rightarrow 0}z\frac{H}{n_{\infty }} =\frac{m \sigma_h \zeta^6}{16}
,\qquad \lim\limits_{z\rightarrow 0}z\frac{D}{%
n_{\infty }}=\frac{3m \zeta^4}{16},
\end{eqnarray}
and finally use  Eq.~(\ref{Eq:ObsHorizon2}) to obtain
\begin{eqnarray}
\lim\limits_{z\rightarrow 0}z\frac{\rho _{h}}{%
n_{\infty }} &=& \frac{3m \zeta^4}{16}\left(1+\frac{2}{\sqrt{3}} \right), 
\\
\lim\limits_{z\rightarrow 0}z\frac{p_{rad,h}}{n_{\infty
}} &=& \frac{3m \zeta^4}{16}\left(-1+\frac{2}{\sqrt{3}} \right), 
\\
\lim\limits_{z\rightarrow 0}z\frac{p_{tan,h}}{n_{\infty
}} &=& \frac{3m \zeta^4}{16}.
\end{eqnarray}
With the help of these relations and Eq.~(\ref{Eq:nHninftyz0limit}), we can find the energy density $\rho _{h}$, radial pressure $p_{rad,h}$, and tangential pressure $p_{tan,h}$ at the horizon for the ultra-relativistic regime presented in Eqs.~(\ref{Eq:nHz0},\ref{Eq:pradHz0},\ref{Eq:ptanHz0}).


\section{Validity of the conditions (i)--(iv) for the Reissner-Nordstr\"om and loop quantum black holes}
\label{App:Validity} 

In this appendix, we verify the validity of the conditions (i)--(iv) in Section~\ref{SubSec:GenericBH} for the RN black hole and the LQBH described in Section~\ref{SubSec:Examples}. Introducing $x := r/r_h$ and $\xi = r_c/r_h\in [0,1)$, we can write the relevant metric functions as follows:
\begin{equation}
\sigma = \left( 1 + \frac{\eta}{x} \right)^2,\qquad
\chi = \left( 1 + \frac{\eta}{x} \right)^2\left( 1 - \frac{1}{x} \right)\left( 1 - \frac{\xi}{x} \right),
\end{equation}
where $\eta = 0$ in the RN case, $\eta = \sqrt{\xi}$ in the LQBH case, and $\eta = 0 = \xi$ in the standard Schwarzschild case. The first derivative of $\chi$ is given by
\begin{equation}
\frac{d\chi}{dx} = \frac{\chi}{x} I,\qquad
I:=\frac{-2\eta}{x+\eta} + \frac{1}{x-1} + \frac{\xi}{x-\xi}.
\end{equation}
It is clear from these expressions that $\sigma,\chi\to 1$ as $x\to \infty$ and that $x^2 d\chi/dx\to 1+\xi-2\eta$, such that condition (i) is satisfied with $2M = (1+\xi-2\eta)r_h$. Note also that $(1+\xi-2\eta) > 0$ in both the RN and the LQGB cases, such that $M > 0$, as required.

Next, regarding conditions (ii) and (iii), we note that $\sigma$ is positive and that $\chi > 0$ for all $x > 1$. Furthermore, the root $x=1$ of $\chi$ is non-degenerate since $\xi < 1$. Next, a simple calculation reveals that
\begin{equation}
x\frac{dI}{dx} = -I - H,\qquad
H := \frac{2\eta^2}{(x+\eta)^2} + \frac{1}{(x-1)^2} + \frac{\xi^2}{(x-\xi)^2},
\label{Eq:D1}
\end{equation}
and since $H > 0$ one concludes from this that the function $x I$ has negative derivative, such that $x I\geq \lim_{x\to\infty} (x I) = 1 + \xi - 2\eta > 0$ for all $x > 1$. In particular, $I$ is positive and (as a consequence of Eq.~(\ref{Eq:D1})) monotonically decreasing for all $x > 1$. This in turn implies that the first derivative of $\chi$ is positive for all $x\geq 1$ and thus the conditions (ii) and (iii) are satisfied. Additionally, the monotonicity of the function $I$ shows that there is a unique photon sphere, whose location $x=x_{ph}$ is determined by the condition $I = 2$, see Eq.~(\ref{Eq:PhotonSphere}).

Finally, we verify condition (iv) which is more involved. For this, we need to demonstrate that the function
\begin{equation}
\frac{\chi ^{\prime \prime}(r)}{\chi^\prime(r)}-2\frac{\chi^\prime(r)}{\chi(r)}+\frac{3}{r}
\end{equation}
has a single zero in the interval $r_h < r < \infty$. Rewritten in terms of dimensionless variables and multiplying by $x I$, this is equivalent to proving that the function
\begin{equation}
G := x\frac{dI}{dx} + I(2-I),
\end{equation}
has a single root in the interval $1 < x < \infty$. Using Eq.~(\ref{Eq:D1}) one can rewrite $G = I(1-I) - H$, which shows that $G < 0$ is negative for $I > 1$ (and in particular $G < 0$ for $1 < x \leq x_{ph}$ where $I\geq 2$). Furthermore, since $I$ and $H$ decay as $1/x$ and $1/x^2$ for $x\to\infty$, it follows that $\lim_{x\to \infty}(xG) = \lim_{x\to \infty}(xI) = 1 + \xi - 2\eta > 0$ which shows that $G$ is positive for sufficiently large values of $x$. In particular, it follows by the continuity of $G$ that it must have (at least) one zero on the interval $x_{ph} < x < \infty$, which is  characterized by the relation
\begin{equation}
I(1-I) = H.
\label{Eq:DRoot}
\end{equation}

It remains to prove the uniqueness of the root of $G$. For this, we differentiate $G$ with respect to $x$. Using Eq.~(\ref{Eq:D1}) again one obtains
\begin{equation}
x\frac{dG}{dx} = -(1 - 2I)(I + H) + 2H + 2J,
\label{Eq:DGx}
\end{equation}
where we have used the relation
\begin{equation}
-x\frac{dH}{dx} = 2H + 2J,\qquad
J := -\frac{2\eta^3}{(x+\eta)^3} + \frac{1}{(x-1)^3} + \frac{\xi^3}{(x-\xi)^3}.
\label{Eq:D2}
\end{equation}
An argument similar to the one used to show the positivity of $I$ reveals that $J > 0$ for all $x > 1$. Now, let $x > 1$ be a zero of $G$. Then, Eq.~(\ref{Eq:DGx}) implies that
\begin{equation}
x\frac{dG}{dx} = -I(1-2I)(2-I) + 2I(1-I) + 2J = I^2(3-2I) + 2J.
\end{equation}
Since $0 < I < 1$ and $J >  0$ at $x$, it follows that $G$ has positive slope at any of its roots. Therefore, the root of $G$ must be unique and describes the location of the ISCO.

More information regarding the dimensionless radii $x_{ph}$ and $x_{ISCO}$, respectively, characterizing the locations of the photon sphere and the ISCO can be obtained from the analysis above. First, it follows from Eq.~(\ref{Eq:DRoot}) that $H\leq 1/4$ at a root of $G$ since $I(1-I)\leq 1/4$ for all $I$. However, as a consequence of the definition of $H$ in Eq.~(\ref{Eq:D1}), one has $H > 1/4$ for all $1 < x < 3$. We conclude from this that $G < 0$ for all such $x$ and consequently, $x_{ISCO} \geq 3$ in both the RN and the LQBH cases. In the RN case, the equations $I=2$ and $G=0$ lead to
\begin{equation}
2x_{ph}^2 - 3(1 + \xi) x_{ph} + 4\xi = 0,\qquad
x_{ISCO}^3 - 3(1+\xi) x_{ISCO}^2 + 9\xi x_{ISCO} - \frac{8\xi^2}{1+\xi} = 0,
\end{equation}
and for small values of $\xi$ one obtains $x_{ph} = 3/2 + \xi/6 + 4\xi^2/27 + \mathcal{O}(\xi^3)$ and $x_{ISCO} =3 + 8\xi^2/9 + \mathcal{O}\left(\xi^4\right)$, whereas in the extremal limit $\xi=1$ one obtains $x_{ph} = 2$ and $x_{ISCO} = 4$. For the LQBH one finds, for small values of $\eta$,
\begin{equation}
I=\frac{1}{x-1}-\frac{2 \eta }{x}+\frac{\eta ^2 (x+2)}{x^2}+\mathcal{O}\left(\eta ^3\right),
\qquad
G = \frac{x-3}{(x-1)^2} - \frac{2\eta(x-3)}{x(x-1)} + \frac{\eta^2(x-7)}{x(x-1)} + \mathcal{O}(\eta^3),
\end{equation}
which implies $x_{ph} = 3/2 - \eta/3 + 29\eta^2/54 + \mathcal{O}(\eta^3)$ and
$x_{ISCO} = 3 + 8\eta^2/3 + \mathcal{O}(\eta^3)$. In the extremal limit $\eta\to 1$, the functions $I$ and $G$ converge pointwise to $4/(x^2-1)$ and $-24/(x^2-1)^2 < 0$, respectively. Therefore, one finds $x_{ph}\to \sqrt{3}$ and $x_{ISCO}\to \infty$, although this case lies outside the physically relevant regime $0 < \eta\ll 1$.


\section{Computation of the critical angular momentum}
\label{App:Lc} 

In this appendix, we provide some details and analytic expressions for obtaining the critical total angular momentum $L_c$. First, we recall that Eq.~(\ref{Eq:Ltilde}) gives the total angular momentum $\tilde{L}(r)$ for which the effective potential has a maximum at a given radius $r$. By substituting this relation in $V_{m,L}(r) = E^2$ one obtains
\begin{equation}
\frac{2m^2\chi(r)^2}{2\chi(r) - r\chi'(r)} = E^2,
\qquad
r_{ph} < r < r_{ISCO},
\label{Eq:Erm}
\end{equation}
which allows one to determine $r$ as a function of $E$. By construction, the corresponding solution $r = r_m(E)$ determines the location of the maximum of the potential when this maximum is $E^2$. Then, $L_c(E)$ is obtained by substituting $r = r_m(E)$ in the expression for $\tilde{L}(r)$ in Eq.~(\ref{Eq:Ltilde}).

For the particular case $E=m$, Eq.~(\ref{Eq:Erm}) reduces to
\begin{equation}
2\chi(r)-2+r\frac{\chi^\prime(r)}{\chi(r)}=0,
\end{equation}
which leads to a seventh order equation for the LQBH case which should be solved numerically. However, for the RN black hole case, it is of the third order with the following solution
\begin{equation}
r_m(E=m)=\frac{4}{3} \left\{M+\sqrt{4 M^2-3 Q^2} \cos \left[\frac{1}{3} \cos ^{-1}\left(\frac{128 M^4-144 M^2 Q^2+27 Q^4}{16 M \left(4 M^2-3 Q^2\right)^{3/2}}\right)\right]\right\},
\end{equation}
and using Eq.~(\ref{Eq:Ltilde}) gives
\begin{equation}
L_c(m) = \frac{m r_m(m)\sqrt{Mr_m(m)-Q^2}}{\sqrt{2Q^2+r_m(m)[r_m(m)-3M]}}.
\end{equation}
In the Schwarzschild limit $Q\to 0$, one obtains $r_m(E=m) = 4M$ and $L_c(m) = 4Mm$, as expected.

\bibliographystyle{unsrt}
\bibliography{refs_kinetic}

\end{document}